\documentclass{aa}
\usepackage[varg]{txfonts}
\usepackage{graphicx}
\usepackage{psfig}	
\usepackage{color}
\usepackage{txfonts}	

\begin{document}

\title{Time variations of narrow absorption lines in high resolution quasar 
spectra\thanks{Based on observations with UVES on
the Very Large Telescope at the European Southern Observatory (under
programs 67.C-0157, 68.A-0170, 074.B-0358, 278.A-5048, 082.A-0569,
087.A-0597 and 087.A-0778) and with HIRES on the Keck Telescope.} }
\titlerunning{Absorption line variations in quasar spectra} 

\author{P. Boiss\'e\inst{1,2}, J. Bergeron\inst{1,2}, J. X. Prochaska\inst{3}, C. P\'eroux
\inst{4}, D. G. York\inst{5}}

\institute{Sorbonne Universit\'es, UPMC Univ. Paris 06, UMR 7095, Institut d'Astrophysique de Paris, 
F-75014, Paris, France\\
           \email{boisse@iap.fr}
       \and
       CNRS, UMR 7095, Institut d'Astrophysique de Paris, F-75014, Paris, France
       \and
       UCO/Lick Observatory, UC Santa Cruz, Santa Cruz, CA 95064, USA
       \and
       Aix Marseille Universit\'e, CNRS, LAM (Laboratoire d’Astrophysique de Marseille) UMR 7326, 
       F-13388 Marseille, France
       \and
       Department of Astronomy, University of Chicago, 5640 South Ellis Avenue, Chicago, 
       IL 60637, USA} 
\authorrunning{P. Boiss\'e et al.}
\date{Accepted for publication : 28 May 2015}

\abstract
{} %{Text of aims} 
{We have searched for temporal variations of narrow absorption lines in high resolution 
quasar spectra. A sample of five distant sources have been assembled, for which 
two spectra - either VLT/UVES or Keck/HIRES - taken several years apart are available.}
%{Text of methods}
{We first investigate under which conditions variations in absorption line profiles can be 
detected reliably from high resolution spectra, and discuss the implications of changes in terms 
of small-scale structure within the intervening gas or intrinsic origin. The targets selected 
allow us to investigate the time behavior of a broad variety of absorption line systems, sampling 
diverse environments: the vicinity of active nuclei, galaxy halos, molecular-rich galaxy disks 
associated with damped Ly$\alpha$ systems, as well as neutral gas within our own Galaxy.}
%{Text of results}
{ Intervening absorption lines from Mg\,{\sc ii}, Fe\,{\sc ii} or proxy species with lines of lower
opacity tracing the same kind of (moderately ionised) gas appear in general to be remarkably stable 
(1$\sigma$ upper limits as low as 10\% for some components on scales in the range 10 - 100~au), 
even for systems at $z_{abs} \approx z_{e}$. Marginal variations are observed for Mg\,{\sc ii} 
lines toward PKS~1229$-$021 at $z_{abs} = 0.83032$; however, we detect no systems displaying changes 
as large as those reported in low resolution SDSS spectra. The lack of clear variations for low $\beta$ 
Mg\,{\sc ii} systems does not support the existence of a specific population of absorbers 
made of swept up gas toward blazars. In neutral or diffuse molecular media, clear changes are seen 
for Galactic Na\,{\sc i} lines toward PKS~1229$-$02 (decrease of $N$ by a factor of four for one 
of the five components over 9.7 yr), corresponding to structure at a scale of about 35~au, in good 
agreement with known properties of the Galactic interstellar medium. Tentative variations are 
detected for H$_2$ $J=3$ lines toward 
FBQS~J2340$-$0053 at $z_{abs} =2.05454$ ($\simeq 35$\% change in column density, $N$, over 0.7 yr in the 
rest-frame), suggesting the existence of structure at the 10 au-scale for this warm molecular gas. 
A marginal change is also seen in C\,{\sc i} from another velocity component of the same absorption 
system ($\simeq 70$\% variation in $N$(C\,{\sc i}).}
{}


\maketitle

\section{Introduction}

Absorption lines in distant active galactic nuclei (AGN) spectra have long been 
recognized as an invaluable source of information on diffuse material lying all along 
these lines of sight (LoS), from local interstellar material within the Milky Way, 
disks or halos of intervening galaxies, intergalactic clouds, up to gas located in the 
vicinity of the AGN itself. As illustrated by numerous studies (see e.g. Noterdaeme 
et al. 2007; Neeleman et al. 2015) many properties of the gas can be investigated 
through a detailed analysis of the absorption 
lines, including ionization level, chemical composition, density and temperature. 

Aside from the broad absorption lines (BAL) 
and intrinsic narrow absorption lines which often display time variations over timescales 
of months or years in the absorber's frame (Hamann et al. 2008, 2011; Filiz Ak et al. 2013 and 
references therein; Misawa et al. 2014; Grier et al. 2015) intervening absorption 
systems observed in quasar spectra are thought to be 
essentially stable in time. However, a recent study based on multi-epoch low resolution SDSS 
spectra (Hacker et al. 2013) suggests that some narrow intervening systems (mainly Mg\,{\sc ii} and  
Fe\,{\sc ii}) might also display variability. These authors propose that time changes are 
due to the transverse motion of the LoS through the absorber coupled to the 
presence of structure in the absorbing gas at scales in the range 10 - 100 au, similar to that
observed for diffuse neutral gas in our own Galaxy (Crawford 2003; Lauroesch 2007, 
Welty 2007, Meyer et al. 2012). 
Indeed, transverse peculiar velocity values of a few 100~km~s$^{-1}$ are expected for the 
target, the observer and the intervening gas, which imply drifts over a few years falling in 
this range (recall that 1 au~yr$^{-1}$ is equivalent to 4.67 km s$^{-1}$). 

Information on spatial structure within the low-ionization medium similar to the one probed 
by multi-epoch SDSS spectra remains quite limited. Observations of 
gravitationally lensed quasars with multiple images have been used to investigate spatial 
variations over scales of a few 100~pc. 
In particular, Rauch et al. (2002) studied the behavior of Mg\,{\sc ii} - Fe\,{\sc ii} 
systems along three adjacent LoS toward Q2237$+$0305 and by comparing the line profiles, 
infer a lower limit of about 0.5 kpc for the overall absorber 
extent and a typical size of the order of 100~pc for cloudlets associated with 
individual velocity components. Similarly, toward the lensed QSO pair APM~08279+5255, 
Kobayashi et al. (2002) derived an estimate
for the cloudlet size of 200~pc, in good agreement with values inferred from modelling
of these absorbers (Churchill \& Charlton 1999; Crighton et al. 2015). 
Such observations are restricted to a 
few systems and do not enable a study of the structure at smaller scales. In our own 
Galaxy, structure in the low-ionization medium has been investigated 
by Welty (2007) who found no change in O\,{\sc i}, Si\,{\sc ii}, S\,{\sc ii} and 
Fe\,{\sc ii} lines over tens of au. Thus, the limited information available 
(see also the recent study by Neeleman et al. 2015) 
does not suggest the presence of significant structure over scales as small as 100~au in the 
low-ionization medium, comparable to the one seen in the neutral medium. 

The results obtained by Hacker et al. (2013) then appear somewhat puzzling and require 
additional investigations. Their study is based on low resolution SDSS spectra which can only 
provide equivalent width measurements and do not give access to the true line profiles 
and to the identification of the individual absorption velocity components. 
If some absorption lines do vary, changes most probably affect only some velocity components 
and should be much more apparent in high resolution spectra. High resolution line 
profiles would also be very useful to assess the reality of any variation detection.
Indeed, when spectral profiles are resolved (or at least partially resolved), 
changes in a specific transition from a given 
species imply a corresponding change for other transitions from the same species that can be 
computed a priori, providing a stringent reliability test. Furthermore, high resolution data 
will minimize the impact of various artefacts like imperfect sky subtraction, bad pixels etc. 
 
Good signal-to-noise (S/N) high resolution quasar spectra have been taken routinely over the 
last 15 years (for instance, in the context of Large Programs conducted on 8 - 10m class 
telescopes; see e. g. Bergeron et al. 2004 and Molaro et al. 2013) 
and for some distant sources, several observations have been performed over time 
intervals reaching ten years. We have thus gathered a sample of five sources with at 
least two 
VLT/UVES or Keck/HIRES spectra taken several years apart. In one case, we acquired a spectrum 
specifically for this study. These lines of sight altogether include a broad variety of 
systems, probing various environments ranging from neutral Galactic gas, 
galaxy disks at intermediate redshifts (i.e. damped Ly$\alpha$ system), 
Mg\,{\sc ii}-Fe\,{\sc ii} systems due to intervening galaxy halos as well as systems 
with a redshift close to the emission redshift. Several of our targets fall in the blazar 
category, which will allow us to investigate the possible presence of narrow systems with a
large ejection velocity, in line with the scenario put forward by Bergeron et al. (2011).

The purpose of our study is to explore the potential of multi-epoch high-resolution 
observations of AGN to investigate the small-scale structure in intervening gas and 
possible intrinsic variations for narrow low ejection velocity systems. 
In Sect. 2, we describe the method (expected LoS drifts, selection of 
transitions which are best suited for a search of variations, etc) while the data and 
raw results are presented in Sect. 3. Their implications are discussed in Sect. 4. 
We summarize our results and mention some prospects for further variability 
studies in Sect. 5.

%%%%%%%%% Sect. 2
\section{Detecting absorption line variations}

When considering two spectra of the same target taken over some time interval, an important 
parameter is the drift of the LoS relative to a foreground absorber. 
We shall first discuss how one can estimate this scalelength over which the comparison of 
the two observations will provide structure information.
Next, we examine which line parameters (the opacity in particular) yield the best 
sensitivity to a given fractional column density change.  
Finally, we discuss methods which can be used to reliably detect line variations and assess 
their significance.
 
\subsection {Line of sight drifts relative to the absorber}

\begin{figure}
   \centering	
    \includegraphics[width=9.cm,angle=0]{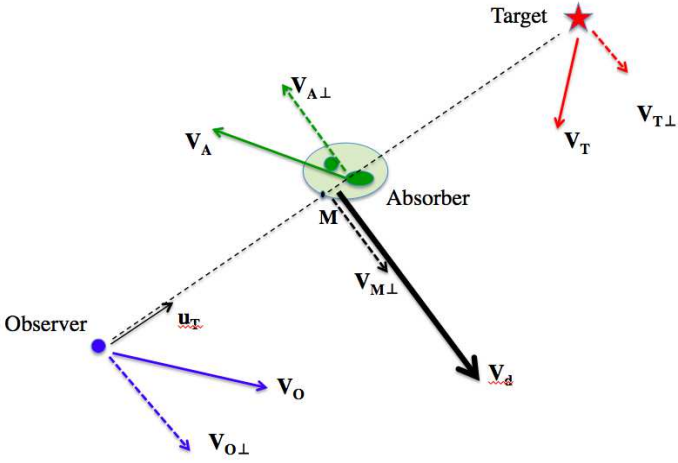}  
    \caption{The observer and target move with 3D velocities
    $\mathbf{V_{O}}$ and $\mathbf{V_{T}}$ in the cosmological frame. 
    Their transverse component
    $\mathbf{V_{O\,\perp}}$ and $\mathbf{V_{T\,\perp}}$ determines the transverse
    velocity $\mathbf{V_{M\,\perp}}$ of the impact point M. The latter combines with
    the transverse velocity of the absorber $\mathbf{V_{A\,\perp}}$ to define the relative
   drift velocity of the LoS with respect to the gas, $\mathbf{V_{d}}$, which sets the 
    scale probed within the absorber during a given time interval. }
     \label{drawing}	
   \end{figure}

For now, we consider that the target is point-like and defines an ``ideal'' LoS, 
intersecting an intermediate redshift absorber at point M (hereafter the ``impact point''; 
Fig. \ref{drawing}). As mentioned above, it is the 
combination of small-scale density or velocity structure
in the intervening gas and the transverse motion of the LoS through this material
which can potentially induce line variations. 
The appropriate frame for computing the velocities of interest 
is a cosmological one, in which the  Cosmic Microwave Background (CMB) is globally isotropic. 
The motion of both the target and the observer determines the drift of the LoS in this frame. 
The observer's motion is due mainly to the peculiar motion of the Milky Way and to Galactic 
rotation. The combination of the two corresponding velocities is directly constrained by the 
detection of the dipolar anisotropy in the CMB radiation indicating that
the Solar System is moving at 368 $\pm$ 2 km s$^{-1}$ relative to the observable Universe 
toward the direction $l_o=263.85 \deg$ and $b_o=48.25\deg$ 
(Fixsen et al. 1996). 

Let us denote by $\mathbf{V_O}$ the corresponding velocity vector. Consider
a target $T$ with Galactic coordinates ($l_T, b_T$) and let $\mathbf{u_T}$ be the 
unit vector pointing in this direction (Fig. \ref{drawing}). 
We are concerned only by the transverse component, 
thus by the projection, $\mathbf{V_{O\,\perp}}$, of $\mathbf{V_O}$ 
onto a plane normal to $\mathbf{u_T}$.\\
This projected velocity can be written as
\begin{equation}
% equation 1
\mathbf{V_{O\,\perp} = V_O -
(u_T \,.\,V_O)\,u_T} = V_{O}\, \mathbf{
\left(u_O - (u_T\,.\,u_O)
\,u_T\right)},
\end{equation}
where $\mathbf{V_{O}} = V_{O} \,\mathbf{.\,u_O}$. From these relations, 
and using the components ($\cos b_T \cos l_T$, $\cos b_T\sin l_T$, $\sin b_T$) of the unit vector 
associated with Galactic coordinates $(l_T, b_T)$, it is 
straightforward to compute $V_{O,\,\perp}$ for any specific target (for the five targets listed 
in Table 1, $V_{O,\,\perp} = 288, 368, 129, 365$ and 12~km s$^{-1}$ respectively). 
In the most favorable case for which $V_{O,\,\perp} = V_{O}$,  this motion alone results in 
a shift of about $4 \times 10^{-3}$~pc or 800~au near the observer, over a time interval of 
10 years. 

%__________________________________________________%    	
%
%-----------------------------------------------------------

Unfortunately, we cannot estimate the transverse velocity of the target 
because the proper motion onto the sky is much too small to be measurable, contrary 
to interstellar medium studies, in which the proper motion of nearby bright stars is known. 
Unless the target belongs to a peculiar 
environment like a galaxy cluster or experienced an ejection event, the magnitude of its 
peculiar velocity should be also of the order of a few 100 km s$^{-1}$. The transverse velocity 
of the impact point M, can then be written as, 
\begin{equation}
% equation 2
\mathbf{V_{M\,\perp}} = \alpha \;\mathbf{V_{T\,\perp}} + 
(1-\alpha)\; \mathbf{V_{O\,\perp}},
\end{equation}
where $\alpha = d_{A}/d_{T}$, $d_{A}$ being the distance to the absorber and 
$d_{T}$ that to the target. 
% ({\it comment on how one should adapt this formula to cosmology ...}.) 
Finally, one must take into account the velocity $\mathbf{V_A}$ of the 
absorber in the cosmological frame or rather its transverse component, 
$\mathbf{V_{A\,\perp}}$. In the end, the value of interest is the relative 
drift velocity 
\begin{equation}
% equation 3
\mathbf{V_{d}} = \mathbf{V_{M\,\perp} - V_{A\,\perp}}.
\end{equation} 
Again, $\mathbf{V_{A\,\perp}}$ cannot be determined (interstellar studies suffer from 
the same difficulty) and only a typical value of a few hundreds of km s$^{-1}$ can be 
considered (as with the observer, the actual value results from the combination of the 
bulk intervening galaxy motion and of the absorbing gas internal kinematics, e.g. 
rotation, infall or outflow). A maximum value for the drift velocity, $V_d$, can be 
estimated assuming an optimal configuration in which  
$\mathbf{V_{O\,\perp}}$ and $\mathbf{V_{T\,\perp}}$ are parallel while 
$\mathbf{V_{A\,\perp}}$ points in the opposite direction. In that case, 
$V_d$ can reach values $V_d= \max \,(V_{O\,\perp}, V_{T\,\perp}) + V_{A\,\perp}$, as large as
about 500~km s$^{-1}$.

Eqs. (2) and (3) are valid in the local 3D space. On large cosmological scales, there is no 
unique way to define distances. Thus, these relations can hold only at low redshifts where 
all cosmological distances become equivalent. Since the exact values of both $V_{T\,\perp}$ and 
$V_{A\,\perp}$ are unknown for any specific target, it is sufficient for our purpose to
estimate drift velocities by i) inserting 
into the above equations typical values for $V_{T\,\perp}$ and $V_{A\,\perp}$, ii) using 
the angular size distance for $d$, and iii) taking into account time dilation
(a time interval $\Delta t$ in the observer's frame corresponds to 
$\Delta t/(1+z_A)$ or $\Delta t/(1+z_T)$ in the absorber's or target's frame respectively, 
which is equivalent to reduce $V_{A\,\perp}$ and $V_{T\,\perp}$ by a factor of $(1+z_A)$ or 
$(1+z_T)$).

In fact, a quasar is an extended continuum source with a 
typical size in the range 10 - 100~au (the size of the accretion disk responsible for the 
optical emission; Dai et al. 2010). Thus, contrary to observations of moving stars used to 
probe interstellar gas structure, significant averaging of the 
foreground structure can occur, implying a reduced amplitude for absorption 
line variations. 
The situation may be different for blazars. 
These sources undergo spectacular photometric variations and, when in a ``high state'', 
their continuum is dominated by emission from the relativistic jet.
Changes in the morphology of the source itself can also occur, which would be equivalent 
to an additional 
contribution to the drift of the LoS.  
Time changes in the 21~cm absorption seen toward AO~0235$+$164 and PKS~1127$-$14 by Wolfe et al 
(1982) and  Kanekar \& Chengalur (2001) are precisely attributed to such an effect.

When dealing with absorption by Galactic interstellar material (four out of our five selected 
targets display relatively strong Ca\,{\sc ii} or Na\,{\sc i} absorption from Milky Way gas), 
the appropriate frame to consider is the Local Standard of Rest (LSR) in which interstellar 
gas is at rest in average. In this case, the drift of the LoS is determined by the peculiar 
motion of the Sun in this frame, $\mathbf{V_S}$, or more specifically, by its 
projection onto a plane perpendicular to the LoS considered. This question is discussed 
in more detail in Boiss\'e et al. (2013).

\subsection {Sensitivity to absorption line variations}

In order to select the absorption lines which are best suited to 
search for time variations, we now quantify the sensitivity 
to a given fractional 
column density ($N$) change. Obviously, very weak or, on the contrary, 
saturated features will  be of little use: there must be some optimal 
intermediate value for the opacity. 

Consider a fully resolved normalised profile, $I_n(v)$, expressed in the velocity scale,  
\begin{equation}
%equation 4
I_n(v)= e^{-\,\tau (v)},
\end{equation}
where $\tau(v)$ is the line opacity at velocity $v$, written as 
\begin{equation}
%equation 5
\tau(v) = k\, \lambda_0 \,f \,N(v),
\end{equation}
where $k$ is a constant, $\lambda_0$ the rest wavelength of the transition, $f$ the oscillator 
strength and $N(v)\, dv$ the column density for species within the velocity interval 
$[v , v+dv ]$. Let us first assume a spectrum noise which is independent on 
$I_n$. The sensitivity can then be expressed as
\begin{equation}
%equation 6
\biggr\vert \frac {\delta I_n}{\delta N/N} \biggr\vert = \tau \,\, e^{-\tau}.
\end{equation}
It reaches its maximum at $\tau = 1$  for which $I_n \approx 0.37$. The interval over which 
the sensitivity is reduced by less than 20\% is $0.61 \le \tau \le 1.53$, 
which corresponds to $0.22 \le I_n \le 0.54$. 

An alternative case of interest is that of photon noise, for which the spectrum r.m.s. 
follows a $\sqrt{I_n}$ law; the ability to detect a given variation $\delta I_n$ is 
then proportional to 
\begin{equation}
\biggr\vert \frac {\delta I_n}{\sqrt{I_n} \; \delta N/N} \biggr\vert = 
\tau \, e^{-\frac{\tau}{2}}.
\end{equation}
This expression reaches its maximum at $\tau = 2$ where $I_n \approx 0.14$ 
(the sensitivity remains good in the sense defined above for 
$1.2 \le \tau \le 3.1$).
In practise, observed spectra are within these two extreme cases; we conclude
that the sensitivity for the detection of variations is optimal for 
{\it intermediate} opacity values (typically 1.5), corresponding to normalised 
intensities of about 0.2. 

One should keep in mind that generally, depending on the instrumental resolution 
and Doppler $b$ parameter value, absorption profiles will  
not be fully resolved, implying apparent opacities lower than the true ones 
(Savage \& Sembach 1991). For instance, 
with an instrumental resolution of 6.6~km s$^{-1}$, the depth of a Gaussian line is reduced 
by a factor of 0.45 for $b=2$~km s$^{-1}$. Then, if the true opacity at 
line center is $\tau_0=1$, the apparent opacity becomes 0.33. Therefore, to 
conclude on the sensitivity of a given line to changes in $N$, one cannot rely only on 
its observed profile. Fortunately, several transitions from the same species 
are often available
with various $f$ values. Simultaneous fitting of the corresponding line profiles
will then allow us to determine the true opacity and assess whether or not these
transitions are appropriate for a variability search.

\subsection {Characterisation of line variations}

Our aim is to detect line profile variations and to quantify their significance. 
A first direct indication can be obtained  by superposing the two successive spectra 
after careful normalisation and adjustment of the wavelength scales if necessary.  
Such a raw comparison is meaningful only if the spectral resolution, $R$, is the same 
at the two epochs. If this is not the case, the higher $R$ spectrum has to be degraded 
to the lower resolution value. 
Fortunately, the HIRES and UVES data used in this paper have very similar nominal 
resolutions - FWHM = 6.25 and 6.6~ km s$^{-1}$ respectively - and we could verify that, 
given the S/N ratio, such a slight difference has a negligible effect on the profile 
of the absorption features discussed in this paper (cf Sect. 3). Differences in the 
line spread function (LSF) might also play a role. One can use unresolved lines to 
check that such an effect does not affect the comparison of successive profiles.

It turns out that for the systems investigated here, the two successive spectra 
available rarely display marked, unambiguously significant differences. Thus, in order 
to i) increase our ability to detect weak variations 
by relying on the presence of several transitions from a given species (e.g. Fe\,{\sc ii}, 
or H$_2$) and ii) take into account differing $R$ values if necessary, a more powerful 
method is to 
simultaneously fit all profiles corresponding to these transitions, at each epoch. 
To this purpose, we used VPFIT\footnote{http://www.ast.cam.ac.uk/~rfc/vpfit.html}
a routine which uses multicomponent Voigt profiles (convolved with the instrument profile, 
which we assume to be Gaussian here) and 
yields the redshift $z_i$, column density $N_i$ and Doppler parameter $b_i$ for each velocity 
component, together with their 1$\sigma$ uncertainties. Comparison of the 
values obtained for each epoch 
then allows us to determine whether some of these parameters have undergone significant time 
changes. In all cases, the same sets of components  were appropriate to fit the profiles at 
both epochs, with no significant change in $z_i$, nor in  $b_i$.  
Some time variations in $N_i$ are  detected or suspected for a few absorptions systems 
(see Sect. 3). 
To assess their significance, we compute the difference between epoch 1 and 2, 
$\Delta N_i = N_{i,2} - N_{i,1}$ and its uncertainty $\sigma (\Delta N_i)$ by combining 
uncertainties on $N_{i,1}$ and $N_{i,2}$ in quadrature. This procedure provides a formal 
significance which properly takes into account the noise level in both spectra. Other 
sources of uncertainty related to continuum placement or to the choice of
velocity components are more difficult to quantify, especially
when variations are small. As a conservative approach, variations with a significance level
above 3.5$\sigma$ and an associated relative column density variation,
$\Delta N /\langle N \rangle$ above 25\% are qualified as "significant". Variations at more
than 2$\sigma$ but which do not fulfill these two criteria are considered
as “marginal”.

When several transitions from the same species are available, the quality of the fit obtained 
provides a stringent test of the internal consistency of the data and gives a powerful means to 
rule out false variations due to instrumental artefacts. Indeed, changes must not only be 
present for all transitions but their magnitude must be consistent with the $f$ values. 
In this regard, species such as Fe\,{\sc ii}, C\,{\sc i}, or H$_2$, which display 
many transitions from a given level within a relatively narrow wavelength interval are 
especially suitable to probe weakly ionized, neutral or molecular gas, respectively.

Often, several species can be used to trace a given phase (e.g. Mg\,{\sc ii}, 
Fe\,{\sc ii}, S\,{\sc ii}  for weakly ionized gas). One can then select among 
all lines detected from these species those i) which opacity falls in the optimal range 
discussed above, and ii) occurring in spectral regions of good S/N ratio.

\begin{table*}[ht]          
\caption[]{The sample.}        
\begin{center}
%\begin{tabular}{lcccccr}        
\begin{tabular}{l@{\hspace{1.4mm}}c@{\hspace{1.5mm}}c@{\hspace{3.6mm}}c@{\hspace{0.5mm}}c@{\hspace{0.5mm}}c@{\hspace{3.0mm}}c@{\hspace{1.mm}}c}
\hline        
\noalign{\smallskip}        
  target & class & coordinates& $z_{\rm em}$ & spectrograph & date & $\lambda$ range
& $\Delta t,n$ \\        
common name  & & J2000 & & & & nm &  \ \ min \   \\        
\noalign{\smallskip}        
\hline        
\noalign{\smallskip}
AO 0235$+$164 $^H$ & BL-$\gamma$ &  023838.9$+$163659  & 0.937 & HIRES & 12-1997 &
367-612 & 40,4   \\
 & & & & UVES & 02-2007  & 303-388$+$58-668 & 60,4    \\
\hline        
PKS 0458$-$020 & FS-$\gamma$ & 050112.8$-$015914 & 2.286 & HIRES & 02-1995 & 394-629
& 480$^a$ \\
 & & & & UVES & 10-2004  & 390-665+$$670-850 & 75,4\\
\hline        
PKS 1229$-$02     & FS-$\gamma$ &  123200.0$-$022404  & 1.044 & UVES & 02-2002 &
326-445$+$458-668 & 60,3    \\
 & & & & UVES & 04-2011  & 326-445$+$458-668 & 50,1    \\
\hline        
PKS 1741$-$038     & FS-$\gamma$ &  174358.8$-$035004  & 1.054 & UVES & 06-2001 &
414-621 & 40,2    \\
 & & & & UVES & 04-2011  & 414-621 & 50,5    \\
\hline        
FBQS~J2340$-$0053 & FS & 234023.7$-$005327 & 2.085 & HIRES & 08-2006 & 306-589  &
250$^a$ \\
 & & & & UVES & 10-2008 & 330-450$+$478-681 & 75,6\\
\noalign{\smallskip}        
\hline        
\noalign{\smallskip}
\multicolumn{8}{l}{class - BL: BL Lac object, FS: Flat Spectrum Radio QSO (FSRQ),
$\gamma$: Gamma-ray emitter. The four $\gamma$-ray emitters are blazars.}\\
\multicolumn{8}{l}{spectrograph - spectral resolution: FWHM(HIRES)=6.25 km s$^{-1}$,
FWHM(UVES)=6.6 km s$^{-1}$.}\\ 
\multicolumn{8}{l}{$\Delta t,n$ \ Exposure time per individual exposure and number
of exposures.}\\
\multicolumn{8}{l}{$^H$ In a very high state at both epochs. The new $z_{\rm em}$ is
derived from 
[Ne\,{\sc iv}] emission.}\\
\multicolumn{8}{l}{$^a$ Total exposure time.} \\
\end{tabular}        
\end{center}        
\label{obs}        
\end{table*}
%      
%%%%%%%%%%% Sect. 3 
\section{Variability study of five blazars with low-moderate $\beta$ systems}

\subsection{The absorber sample}

One of our initial motivations was to explore the link between gas swept by the AGN jet 
and low $\beta$ absorbers, as suggested by Bergeron et al. (2011). We then investigated 
among their blazar sample objects with jets, thus 
radio-loud and $\gamma$-ray emitters (detected by Egret, Integral or Fermi/LAT). 
An additional selection criterion is the existence of Mg\,{\sc ii} absorption systems 
at low relative velocity $\beta\equiv \Delta v(z_{\rm em},z_{abs})/c<0.15$ 
with a rest-frame equivalent width $W_r(2796) >0.3$~\AA, as well as intervening 
% $w_{\rm r}$
systems and 
Damped Ly-$\alpha$ (DLA) systems at high $\beta$. We then searched the VLT-UVES and 
Keck-HIRES archives for high-spectral resolution spectra of this blazar subsample, 
and selected targets observed several years apart, possibly about one decade. 
We checked that the signal-to-noise per pixel was S/N $\gtrsim 10$ for the main 
absorptions of interest: the Mg\,{\sc ii} doublets at $\beta<0.15$ and the Galactic 
Na\,{\sc i} doublets. 

One blazar satisfying the above selection criteria, is the extensively 
studied, highly variable BL Lac object AO 0235$+$164 for which variations of 21 cm intervening 
absorption at $z_{\rm abs}=0.524$ had been reported three decades ago (Wolfe et al. 1982). 
This absorber has been identified as a BAL quasar with very faint extended 
emission (Burbidge et al. 1996).
The variations were interpreted as most probably due to variations in the brightness 
distribution of the background radio source. Another blazar is PKS 1741$-$038, which was 
re-observed by us with UVES in 2011 to search for variability of its two strong  
Mg\,{\sc ii} absorbers at $\beta=0.074$ and 0.287. For both objects, the available UVES 
and HIRES spectra have been obtained at epochs about 9.5 yr apart in the observer's frame.

We also searched for high resolution spectra of PKS 1229$-$02, a blazar for which we 
obtained deep HST imaging showing optical emission associated with knots in the 
radio jet (Le Brun et al. 1997). There is a DLA system at 
$z_{\rm abs}=0.3950$ with associated 21 cm absorption, and several Mg\,{\sc ii} absorbers 
at $\beta\leq 0.15$. The above selection criteria are also met for this source and there 
are two sets of high S/N UVES spectra taken at epochs 9.2 yr apart.  

Finally, there are HIRES and UVES data available for two targets with  C\,{\sc i} 
absorption at 
$z_{\rm abs}\sim 1.5$ with associated 21 cm absorption at high $\beta$ (Kanekar et al. 
2010): PKS 0458$-$020 ($z_{\rm abs}=1.5606$) and FBQS~J2340$-$0053 ($z_{\rm abs}=1.3608$), 
the latter with observations about two years apart. In the radio spectrum  of PKS 0458$-$020, 
variable  21 cm 
absorption on time-scale of months has been detected at $z_{\rm abs}=2.0395$ 
(Kanekar et al. 2014 and references therein). There are also DLA systems at low $\beta\leq 
0.10$ toward PKS~0458$-$020 at $z_{\rm abs}=2.0395$ and FBQS~J2340$-$0053 at $z_{\rm abs}=2.0545$. 
The physical conditions in the low $\beta$ DLA system of FBQS~J2340$-$0053,  
with associated  C\,{\sc i} absorption, have been investigated by Jorgenson et 
al. (2010). The multiple system at $z_{\rm abs}=1.3608$ toward FBQS~J2340$-$0053 has been 
extensively studied by Rahmani et al. (2012) to constrain the variation of 
fundamental constants. 

Information on the sample and the observations is given in Table 1 and that on the 
absorption systems in Table 2. For some UVES data, we used the reduced data sets produced 
internally at ESO, based on the latest calibration pipelines and the best available
calibration data. For the other UVES observations, the spectra were reduced by us or made 
available to us by several colleagues (see the acknowledgements). For HIRES spectra, the 
data sets were reduced by Prochaska's team. All redshifts and velocities considered in 
the following are vacuum heliocentric.

In the analysis presented below, we estimate the variation of $N$ between the two epochs and 
use the combined quadratic errors to evaluate its significance level, as described in Sect. 2.3. 
In case of no variation, 
we give the $1\sigma$ limit on $\Delta N /\langle N \rangle$, restricting this evaluation to 
fairly good S/N, thus meaningful limits: $\Delta N /\langle N \rangle <$ 50\%.

\begin{table*} 
\caption[]{The absorption systems.}	
\begin{center}
\begin{tabular}{l@{\hspace{2.5mm}}r@{\hspace{3.5mm}}r@{\hspace{3.5mm}}r@{\hspace{3.5mm}}c@{\hspace{3.5mm}}c@{\hspace{3.0mm}}c}
\hline	
\noalign{\smallskip}	
target  & $z_{\rm abs}$& $\beta$ &  main & strength & S/N$^a$ & associated low ions   \\
 & & &  doublet     &   &  & present in the 2 spectra   \\        
\noalign{\smallskip}	
\hline	
\noalign{\smallskip}
AO 0235$+$164  & $+0.00002$  &  &  Na\,{\sc i}   & w & 67 - 56  & none   \\
 &  0.52432  & 0.235 & Mg\,{\sc ii}   & VSm & \ 28 - 14$^b$  &  Mg\,{\sc i}, Fe\,{\sc ii},  Mn\,{\sc ii}, 
Ca\,{\sc ii}, Ti\,{\sc ii} \\
%, H\,{\sc i}4341: , region not covered by HIRES % Na\,{\sc i}3303:
 &  0.85142  & 0.045  &  Mg\,{\sc ii}  & w &  62 - 50 & none \\
 &  0.85237  & 0.045  &  Mg\,{\sc ii}  & S &  62 - 50 &  Mg\,{\sc i}, Fe\,{\sc ii}  \\
 &  0.85562  & 0.043  &  Mg\,{\sc ii}  & w &  62 - 50 & none \\
\hline
%PKS 0458$-$020 &   $+0.00008$ &   &  Na\,{\sc i}  & S  & gap - 13  & none      \\ 
PKS 0458$-$020 &  0.89025  & 0.503  &  Mg\,{\sc ii}  & S  & \  11 - 13  &     Fe\,{\sc ii}  \\ 
 &  1.52712  & 0.257  &  Mg\,{\sc ii}  & VSm& 15$^b$ - 5 &   Fe\,{\sc ii}    \\ 
 &  1.56055  & 0.244  &  Mg\,{\sc ii}  & S  & 14$^b$ - 5 &   C\,{\sc i}, Fe\,{\sc ii},  
Zn\,{\sc ii}, Cr\,{\sc ii}, Ni\,{\sc ii} \\ 
 &  2.03956  & 0.078  &  Mg\,{\sc ii}  & S  & 10$^{b'}$ - 5 &   Fe\,{\sc ii}, Zn\,{\sc ii}, Cr\,{\sc ii}, 
Ni\,{\sc ii}, Si\,{\sc ii}  \\ 
%%%%%%%% question: only the associated absorption common to the 2 spectra/epochs ?? 
\hline  
PKS 1229$-$02    &   $ -0.00016$ &   &  Na\,{\sc i}  & w & 52 - 15 &  Ca\,{\sc ii}    \\
 &  0.39458  & 0.364  &  Mg\,{\sc ii}  & VSm  & 44 - 13 &  Mg\,{\sc i}, Fe\,{\sc ii}, Mn\,{\sc ii}, 
Ca\,{\sc ii}, Ti\,{\sc ii}  \\
 &  0.69990  & 0.182  &  Mg\,{\sc ii}  & Vw  & 50 - 13 &  none \\
 &  0.75643  & 0.150  &  Mg\,{\sc ii}  & ws  & 50 - 13 & Mg\,{\sc i}, Fe\,{\sc ii} \\
 &  0.76836  & 0.144  &  Mg\,{\sc ii}  & w  & 50 - 14 & none \\
 &  0.83032  & 0.110  &  Mg\,{\sc ii}  & w  & 53 - 14 & Fe\,{\sc ii} \\
 &  0.83102  & 0.110  &  Mg\,{\sc ii}  & Vw  & 53 - 14 & none \\
 &  0.83128  & 0.109  &  Mg\,{\sc ii}  & w  & 53 - 14 & Fe\,{\sc ii} \\
\hline
PKS 1741$-$038   & $-0.00070$ &   &   Na\,{\sc i} & w & 12 - 32 & none   \\
 & $-0.00033$   &     &   Na\,{\sc i}  & S & 12 - 32  &  DIBs   \\
%CH$^+$4233: \\
  &  0.52739  &  0.288  & Mg\,{\sc ii} & w  & 5 - 13  & none   \\
  &  0.52822  &  0.287  & Mg\,{\sc ii} & S  & 5 - 13  & Mg\,{\sc i}, Fe\,{\sc ii} \\
%Ca\,{\sc ii}  \\
  &  0.67846  &  0.199  & Mg\,{\sc ii} & w  & 8 - 18  & none  \\
  &  0.90799  &  0.074  & Mg\,{\sc ii} & VSm  & 10 - 28 & Mg\,{\sc i}, Fe\,{\sc ii} \\ 
%Mn\,{\sc ii}    \\
\hline 
FBQS~J2340$-$0053 & $-0.00002$ & & Na\,{\sc i}  & S  & 29 - 68 & Ca\,{\sc ii}      \\ 
& 1.36064 & 0.261 & Mg\,{\sc ii}  & VSm  & 30$^b$ - 64 &   C\,{\sc i}, Fe\,{\sc ii}, Zn\,{\sc ii}, 
Cr\,{\sc ii}, Si\,{\sc ii}, Ni\,{\sc ii}   \\ 
& 2.05454 & 0.010 & Fe\,{\sc ii}  & VSm  & \ \ 34$^{b'}$ - 56$^{b'}$ & C\,{\sc i}, C\,{\sc i}$^{\star}$, 
Fe\,{\sc ii}, S\,{\sc ii},  Si\,{\sc ii}   \\ 
%% & & &   &   &  &      \\
\noalign{\smallskip}	
\hline	
\noalign{\smallskip}
\multicolumn{7}{l}{$^a$ : S/N per pixel around the main absorption for the older and recent epoch, 
respectively.}\\
%\multicolumn{7}{l}{$^d$ : at least 2 lines per low ion.} \\
\multicolumn{7}{l}{$^{b,b'}$ : Mg\,{\sc ii} region not covered; we list instead the S/N around 
Fe\,{\sc ii}2374-2382,  Fe\,{\sc ii}1608-1611.} \\
\multicolumn{7}{l}{VSm : very strong, highly multiple  absorption,  $w_{\rm r}$(2796) $ \gtrsim 2.0$ \AA.} \\
\multicolumn{7}{l}{S : strong absorption,  $ w_{\rm r}$(2796 or 5891) $ > 0.3$ \AA.} \\ 
\multicolumn{7}{l}{w : weak absorption,  $0.02 < w_{\rm r}$(2796 or 5891) $ < 0.3$\AA.} \\
\multicolumn{7}{l}{ws : although weak, the Mg\,{\sc ii} doublet is saturated.} \\
\multicolumn{7}{l}{Vw : very weak absorption,  $w_{\rm r}$(2796) $ < 0.02$ \AA.} \\
\end{tabular}	
\end{center}
\label{obs}	
\end{table*}
%	
%--------

\subsection{Tracers of the Interstellar Medium} 

The main tracers of the local Interstellar Medium (ISM) detectable in our spectra are the 
Na\,{\sc i} and Ca\,{\sc ii} doublets, Ca\,{\sc i}$\lambda$4227 as well as 
the CH$^+\lambda$4232 and CH$\lambda$4300 molecular lines. Diffuse interstellar bands (DIBs), 
mainly at 4428, 5780 and 5797 \AA, could also be identified in some UVES spectra. 

%------------------------------------------------------------------------------------------%    	
%   \begin{figure*}[hb]	 or [ht]
   \begin{figure*}
   \centering	
     \includegraphics[width=8.5cm,angle=270]{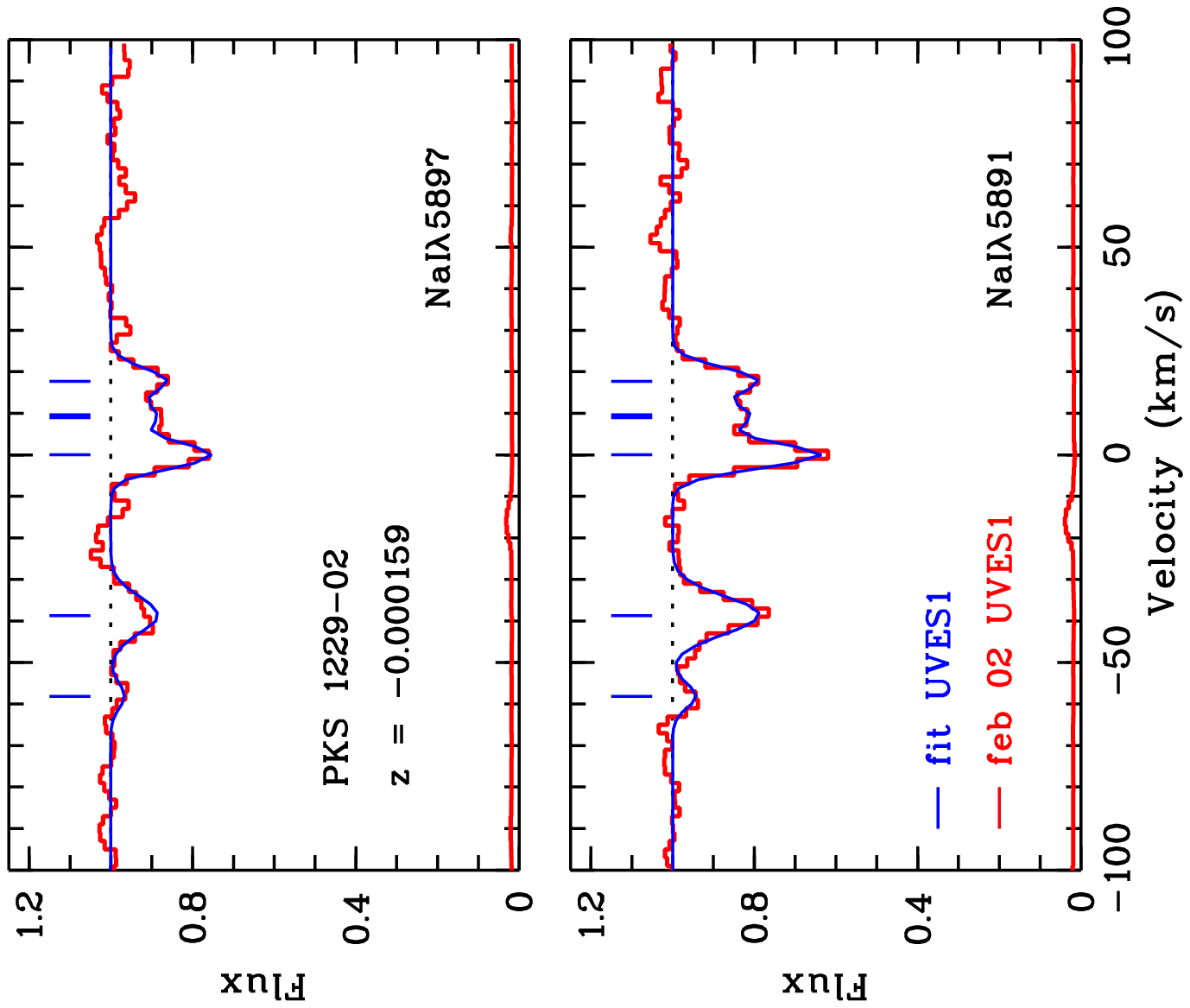}
%\hspace{5mm} 
%\vspace{18mm} 
     \includegraphics[width=8.5cm,angle=270]{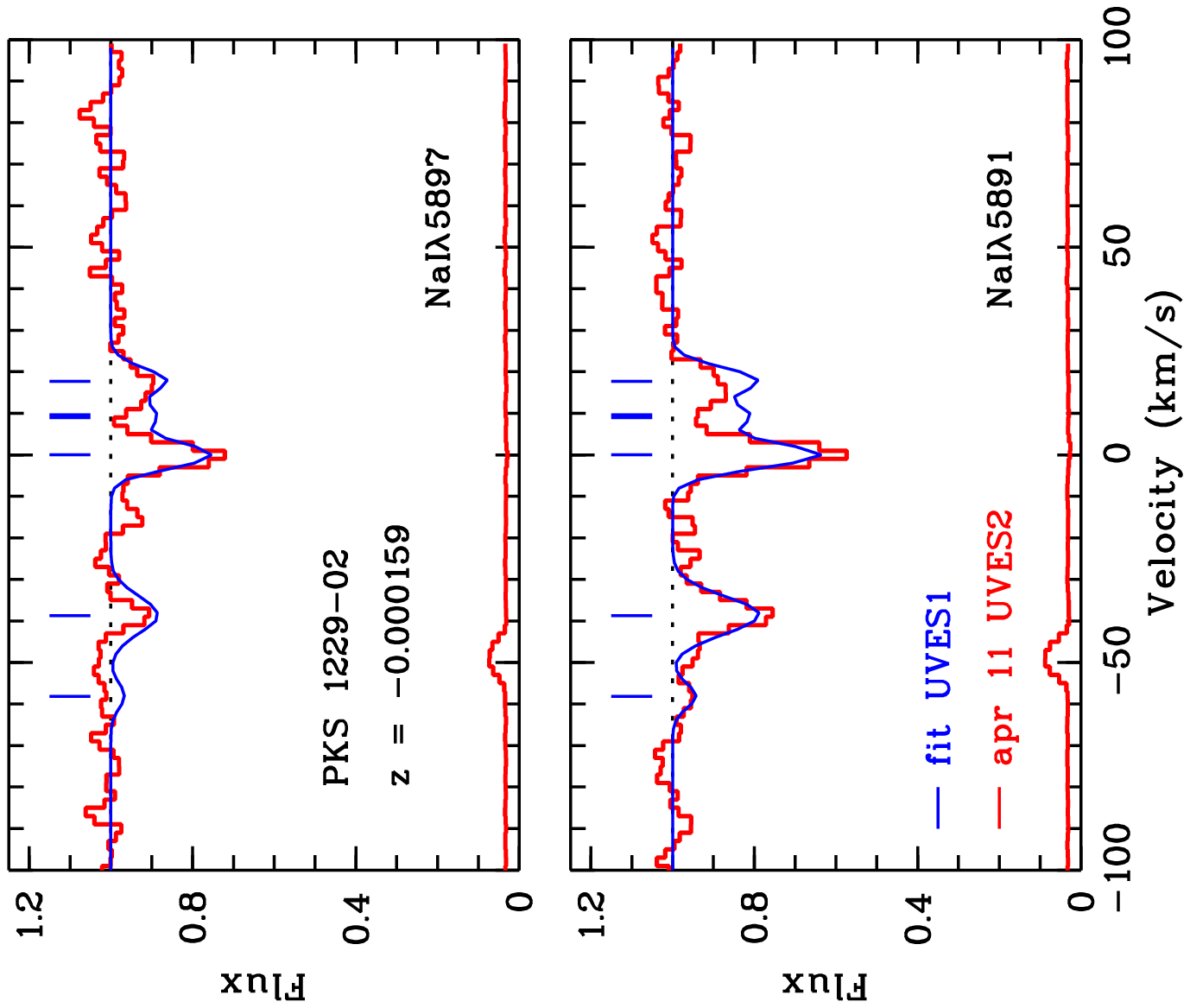}	
%\hspa\includegraphics[width=5.5cm]{NaI0.0vladilo_fitellison.ps}
    \caption{Galactic Na\,{\sc i} absorption toward PKS 1229$-$02: 
    spectrum, its error (red curves) and simultaneous fit to both transitions 
    (blue curve). The UVES 2002 data are shown in 
    the left panel; in the right panel we display the UVES 2011 data together with 
   the fits to the 2002 data (blue). Note the disappearance 
   of the component at  $v_{\rm helio}=9.6$ km s$^{-1}$ (thick vertical tick mark), 
   within a period of 9.2 yr. 
   The heliocentric redshift $z = -0.000159$ (strongest component) corresponds to 
   $v=0$~km~s$^{-1}$.}
   \label{P1229NaI}	
   \end{figure*}	
%-----------------------------------------------------------	

Toward AO 0235$+$164, the Galactic Na\,{\sc i} absorption shows a strong, single component 
at $z_{\rm abs}=+0.00002$. A simultaneous fit to both members of the Na\,{\sc i} doublet yields 
identical values at the two epochs, 9.2 yr apart, within the 1$\sigma$ errors:  
$N$(Na\,{\sc i}) $= (2.51 \pm0.04)\times 10^{12}$ cm$^{-2}$. 

The  Galactic Na\,{\sc i} absorption toward PKS 1229$-$02 is multiple (5 components) and extends 
over 76 km s$^{-1}$. Significant variation is present for the component at 	
$z_{\rm abs}=-0.000127$: in 9.7 yr, it has nearly  disappeared,  
%%%%  or $v_{\rm LSR}\simeq 28$ km s$^{-1}$ 
with $N$(Na\,{\sc i}) decreasing by a factor of four (see Table 3 and Fig. \ref{P1229NaI}). 
The other three 
strong components ($N$(Na\,{\sc i}) $> 10^{11}$ cm$^{-2}$) did not vary ($1\sigma$: 17\%). 

Toward PKS 1741$-$038, Galactic molecular absorptions at $v_{\rm LSR}=2$ km s$^{-1}$ have been 
extensively investigated, with detections of HCO$^+$  (Lucas \& Liszt 1996), CO  (Akeson \& 
Blitz 1999) and C$_2$H (Lucas \& Liszt 2000), together with  variable 21 cm absorption (Lazio 
et al. 2001). The associated Na\,{\sc i} absorption at $z_{\rm abs}=-0.00033$ is multiple (4 
components), partly saturated and blended with atmospheric Na\,{\sc i} emission,  thus 
preventing a search for variations.  
Two well-known DIBs are detected at both epochs, at 5780 and 5797 \AA. Their equivalent 
widths are equal to 0.33 and 0.10 \AA\ respectively, and are similar at two 
epochs 9.8 yr apart, with 
$\Delta W$/$\langle W \rangle = 21$\% $\pm 13$\% and 20\% $\pm 23$\%.
Two other DIBs at 4428 and 6196~\AA, the Ca\,{\sc ii} doublet and CH$^{+}\lambda 4232$ 
are detected but are far too noisy in the 2001 spectrum to investigate variability. 
 
Finally, toward FBQS~J2340$-$0053, the Na\,{\sc i} 5897 line is not covered in the 
HIRES spectrum. The single 
component of the Ca\,{\sc ii} doublet at $z_{\rm abs}=-0.00002$ remains stable 
($1\sigma$: 3.8\%) within 1.9 yr. 

\subsection{Neutral and molecular gas at high $z$} 

\begin{figure*}
   \centering
      \includegraphics[width=10.cm,angle=270]{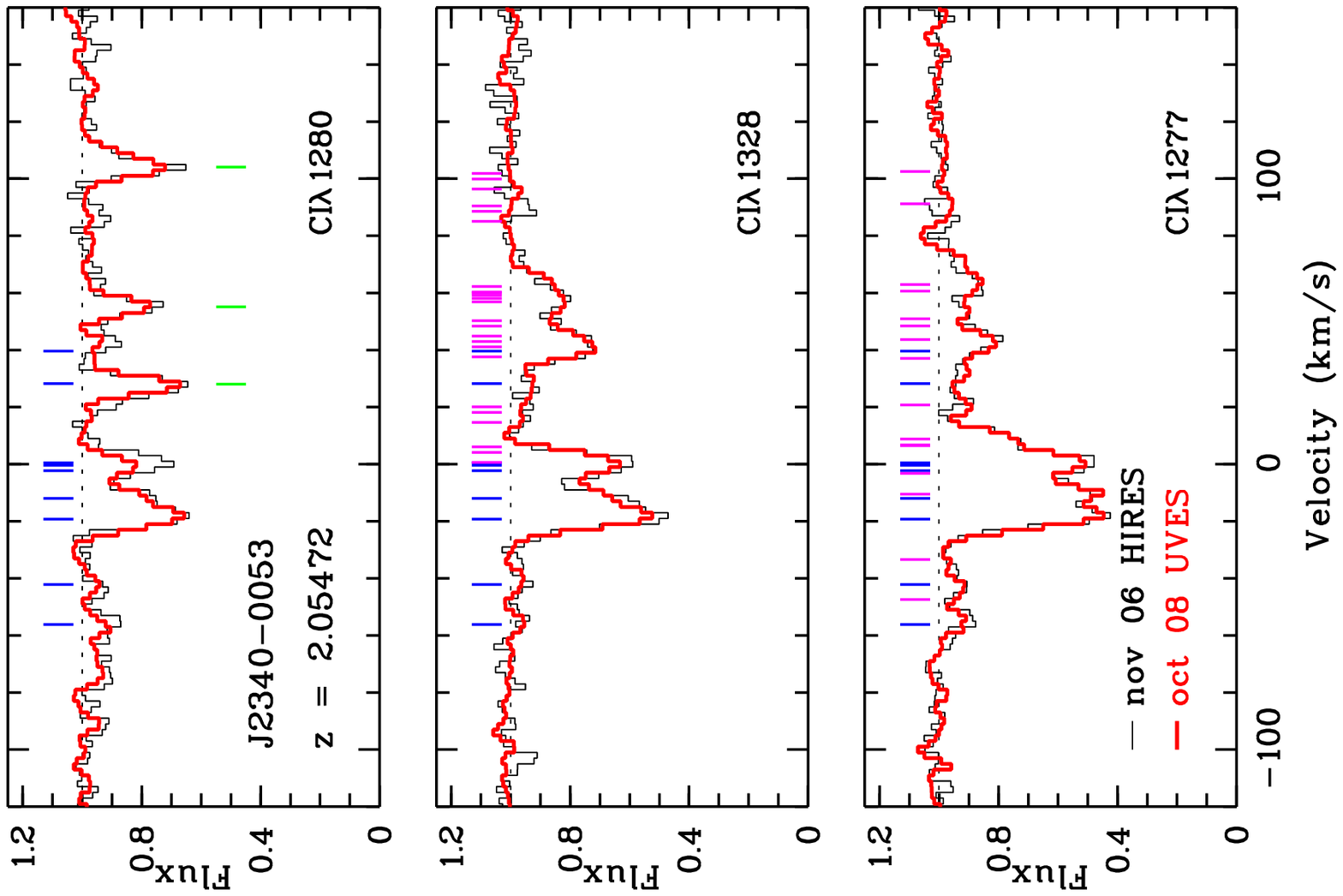}	
\hspace{5mm}
	\includegraphics[width=10.cm,angle=270]{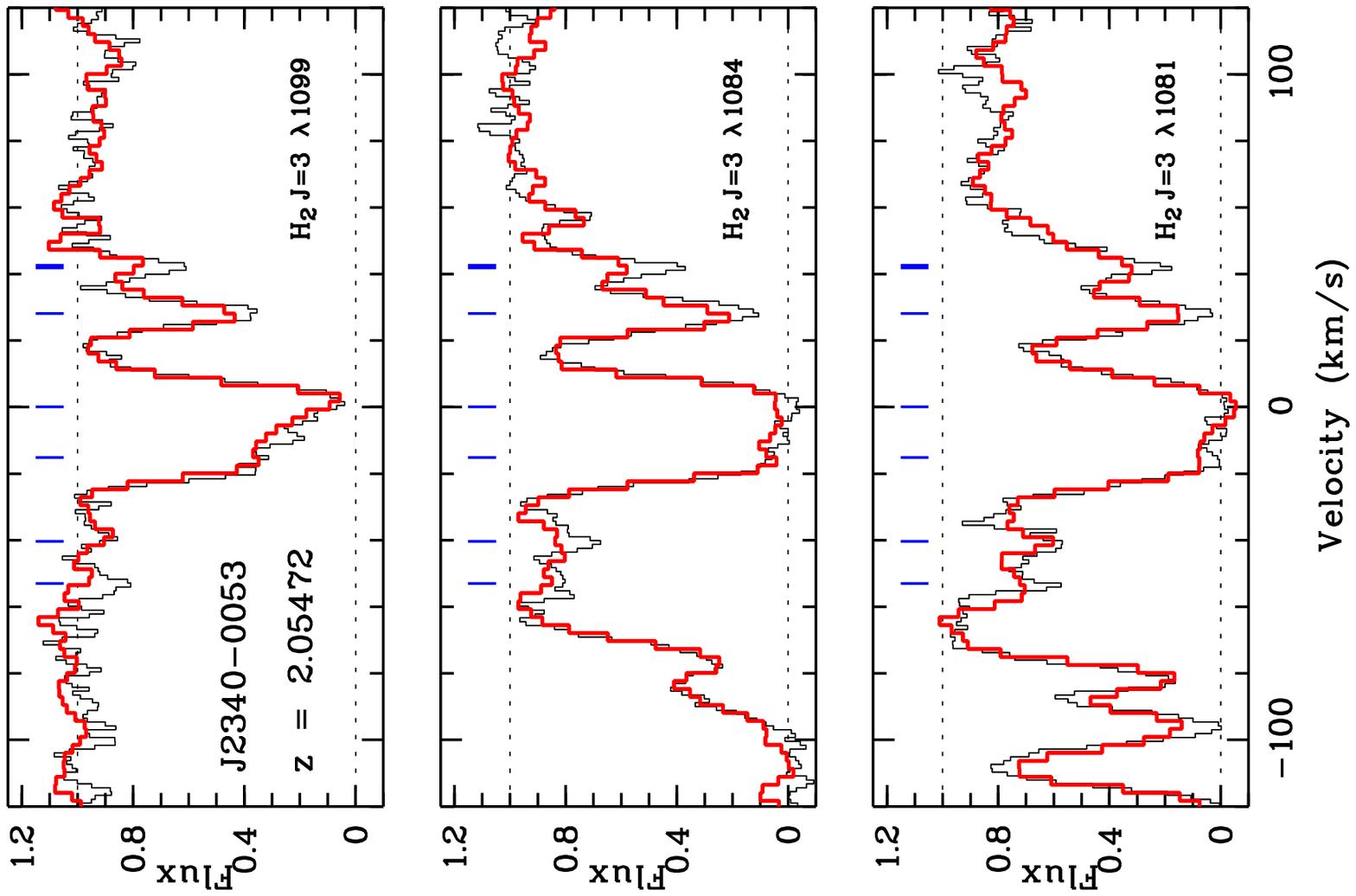}
    \caption{ C\,{\sc i}$\lambda 1277,1280,1328$ and molecular
      H$_2$, $J$=3 $\lambda1081, 1084, 1081$ absorptions at $z_{\rm abs} = 2.05472$ toward  
      FBQS~J2340$-$0053 in the November 2006 (HIRES, black line) and October 2008 (UVES, red line) 
      spectra. 
     {\it Left panel}: the eight C\,{\sc i} components required to fit the profiles are shown by blue 
     tick marks and the marginally varying feature is at $v=0$ (thick tick mark). In the 
     top C\,{\sc i}$\lambda1280$ plot, three C\,{\sc i}$\lambda1656$ lines from the 
     $z_{abs} = 1.3606$ system are also shown (green tick marks).
     For the C\,{\sc i}$\lambda$1277 and C\,{\sc i}$\lambda$1328 transitions, we show the position 
     expected for C\,{\sc i}$^*$ transitions associated with the 8 components (cyan tick marks); note 
     that the features at $v \simeq $ 40 km s$^{-1}$ are mainly due to C\,{\sc i}$^*$ transitions. 
     {\it Right panel}: six H$_2$ $J$=3 components are required to fit the profiles. 
     Marginal variability is seen in the sixth component (thick tick mark) at 
     $v\simeq 40$ km s$^{-1}$ (some difference is also seen in the fifth component but it is 
     significant only at the 2.0$\sigma$ level, instead of 3.1$\sigma$ for the sixth one). 
     For both narrow components displaying marginal variations in C\,{\sc i} ($v\simeq 0$) or H$_2$ 
     ($v\simeq 40$ km s$^{-1}$), we checked that the slightly higher resolution of the HIRES spectrum 
     cannot account for the observed difference between the profiles. 	
}
         \label{CIJ2340}	
   \end{figure*}

Toward PKS 0458$-$020,  C\,{\sc i} is well detected at $z_{\rm abs}=1.5605$ in the HIRES spectrum 
(with a single component, as reported in Kanekar et al. 2010) but of too 
low S/N in the UVES spectrum to get useful variability constraints. 
In FBQS~J2340$-$0053 spectra, C\,{\sc i} is detected at $z_{\rm abs}=1.3606$ 
and 2.0545. The three well separated components of the lower $z$ system are stable 
($1\sigma$: 12\%). The fit of  C\,{\sc i} at $z_{\rm abs}=2.0545$ involves a minimum of 
eight components, two sets being heavily blended (components 3-4 and 5-6 
respectively). Within the blend at $z_{\rm abs}=2.05473$ a very narrow component (number 6) 
shows a possible variation $\Delta N /\langle N \rangle$ of 
69\% in 0.71 yr (rest-frame) at the $3.0\sigma$ significance level (see Table 3 and 
Fig. \ref{CIJ2340} left panel); the $1\sigma$ limits for the other seven components are in the 
range 9-50\%.  

%
%------------------------------------------------------------------------------------------%    	
%   \begin{figure*}[hb]	 or [ht]
%-----------------------------------------------------------	

Molecular hydrogen has already been detected toward FBQS~J2340$-$0053 at $z_{\rm abs}=2.0545$ 
(Jorgenson et al. 2010). The Lyman and Werner transitions display six components (which 
roughly correspond to C\,{\sc i} components number 1, 2, 3-4, 5-6, 7 and 8) and  
there is a possible variation of the sixth H$_2$, $J=3$ weak component at 
$z_{\rm abs}=2.05513$.  
Using the unblended components of the transitions at 1081, 1084 and 1099 \AA\ 
%  1099.79 \AA\ (L1 P3), 1084.56 1081.71 \AA\ (L2 P3) and 1081.71 \AA\ (L2 R3) 
we find  a decrease, $\Delta N /\langle N \rangle = -34$\%, in only 0.71 yr (rest-frame), 
significant at the $3.1\sigma$ level (Table 3  and Fig. \ref{CIJ2340}, right panel).
Unfortunately, the corresponding C\,{\sc i} features are too weak to check whether 
C\,{\sc i} lines display similar variations (note in the left panel of 
Fig. \ref{CIJ2340} that features at $v \simeq 40$ km s$^{-1}$ for the 
C\,{\sc i}$\lambda1277$ and C\,{\sc i}$\lambda1328$ transitions are in fact dominated by 
C\,{\sc i}$^*$ transitions from the strongest components). Conversely, the fourth H$_2$ component which 
corresponds to the second C\,{\sc i} blend (components number 5 and 6) with marginal 
variations is opaque in the available transitions and thus, insensitive 
to time changes. \\

\subsection{The Mg\,{\sc ii}-Fe\,{\sc ii} systems}

\subsubsection{Low $\beta$ Mg\,{\sc ii} systems}

There are only two blazars with low $\beta$ Mg\,{\sc ii} systems of moderate strength. 
For the other three, these systems are either DLAs or extremely strong saturated  
Mg\,{\sc ii} absorbers; we have then searched for proxies of intermediate opacity 
(e.g. Fe\,{\sc ii} lines) to constrain variability. 

%__________________________________________________%    	
   \begin{figure}
   \centering	
    \includegraphics[width=10.cm,angle=270]{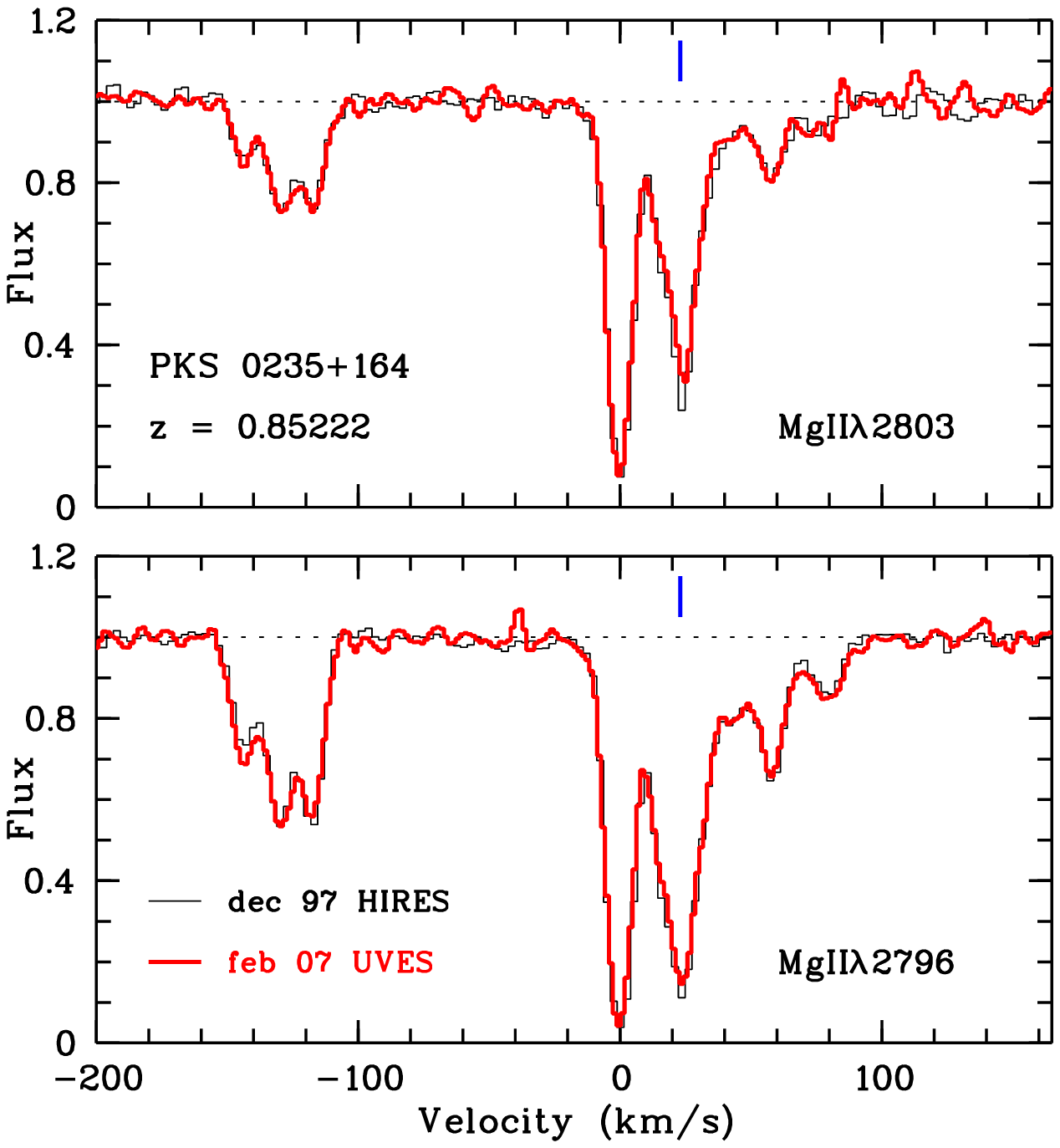}  
    \caption{Intervening Mg\,{\sc ii} absorption at $z_{\rm abs}=0.852$ toward AO 0235$+$164
    in the December 1997 (HIRES, black line) and February 2007 (UVES, red line) spectra
    (lower panel: Mg\,{\sc ii}$\lambda$2796; upper panel: Mg\,{\sc ii}$\lambda$2803).
    The blue vertical tick mark underlines a marginally variable blended component at 
    $z_{\rm abs}=0.85237$, although at a 2.3$\sigma$ significance level only.  
}	
         \label{P0235MgII}	
   \end{figure}
%
%-----------------------------------------------------------

Toward AO 0235$+$164, there are three Mg\,{\sc ii} systems at low $\beta$: 
$z_{\rm abs}=0.8514$ (three fitted components), 0.8523 (seven components) and 
0.85562 (single).
There is no variation of the triple and single systems ($1\sigma$ limits on 
$\Delta N/\langle N \rangle$ of 9.5\% and 6.3\%, respectively).
For the multiple system, one Mg\,{\sc ii} component at $z_{\rm abs}=0.85236$,
blended with two weaker ones at $z_{\rm abs}=0.85231$ and 0.85241, shows a tentative 
variability of its column density $\Delta N /\langle N \rangle$ of 53\% in 
5.0 yr (absorber rest-frame), but at a low significance level of only $2.3\sigma$ 
(see Table 3 and Fig. \ref{P0235MgII}). For these three
components, the associated  Fe\,{\sc ii} doublet is stable ($1\sigma$: 5.2\%)

\begin{figure*}
   \centering	
     \includegraphics[width=8.5cm,angle=270]{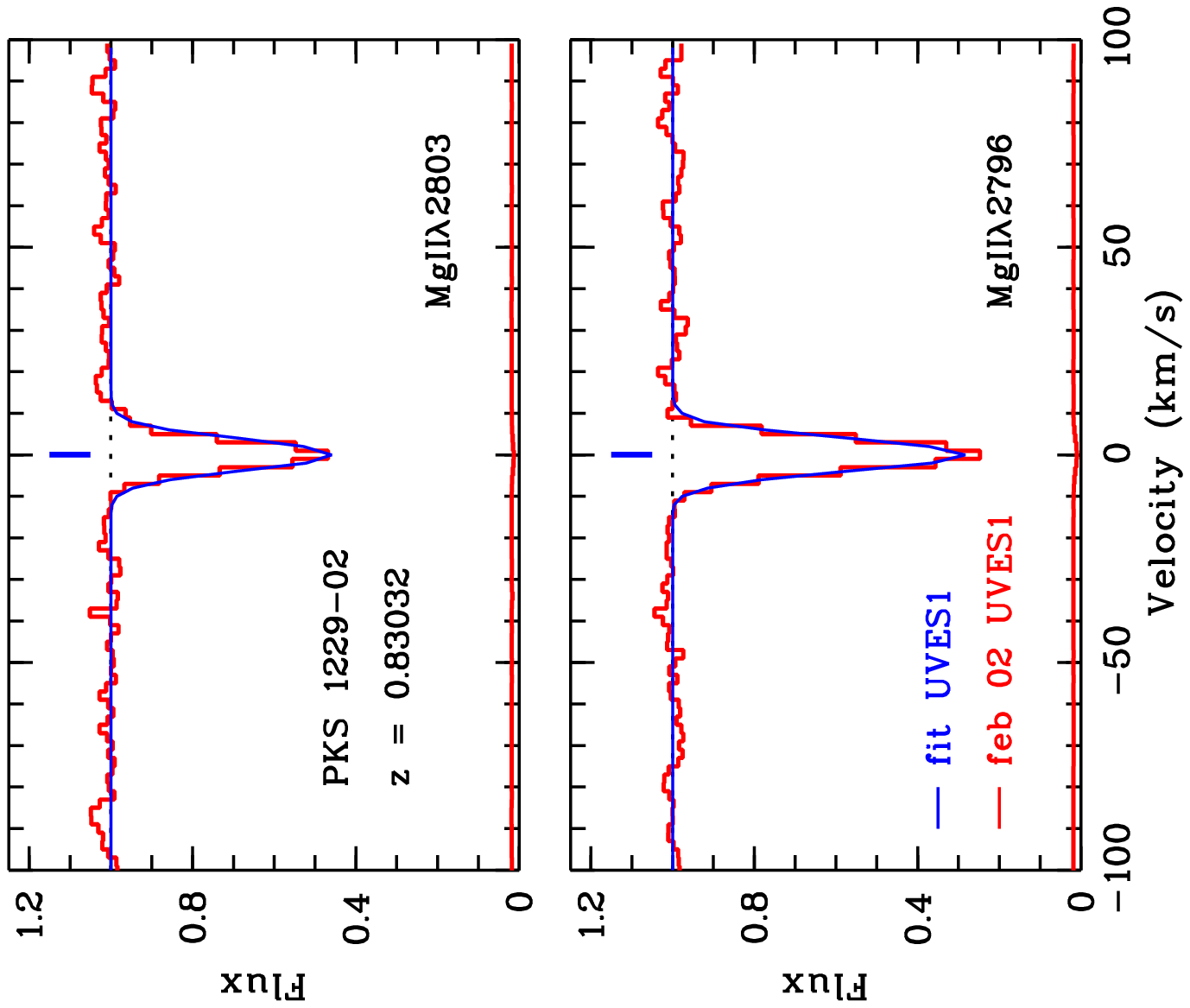}
%%     \vspace{18mm}
     \includegraphics[width=8.5cm,angle=270]{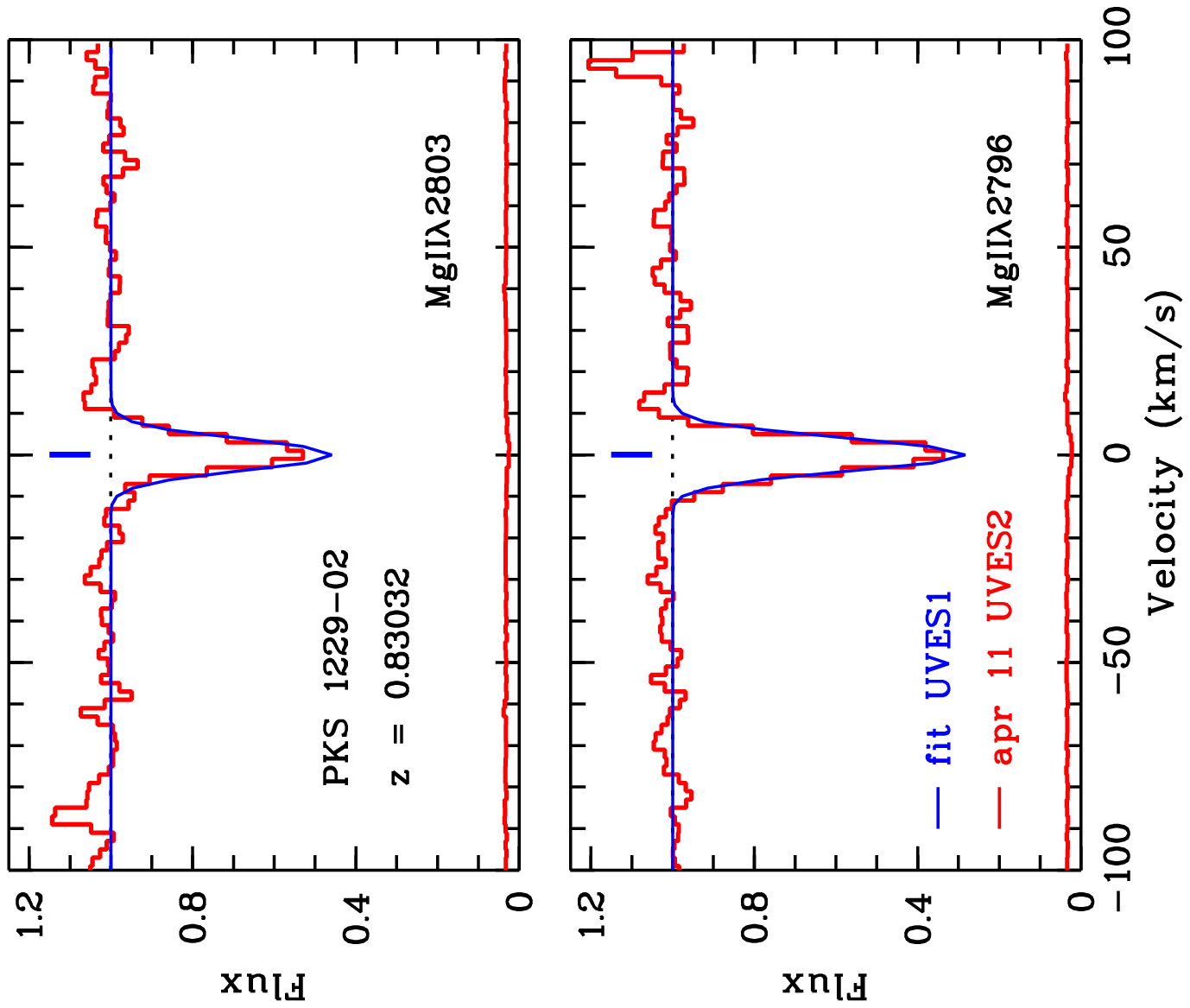}
    \caption{Intervening Mg\,{\sc ii} absorption at $z_{\rm abs}=0.83032$ toward PKS 1229$-$02: 
    spectrum and its  error (red curves) together with simultaneous fit to both transitions 
    (blue curve). 
    In the left panel, we show the UVES 2002 data, 
    while the right panel displays the UVES 2011 data with the fit performed on the 
    UVES 2002 spectrum. 
    The blue vertical tick mark underlines the marginally variable 
    Mg\,{\sc ii} doublet. 
}	
         \label{P1229MgII}	
   \end{figure*}

Toward PKS 1229$-$02, there is a triple, unblended system at $z_{\rm abs}=0.830$. The 
strongest component at $z_{\rm abs}=0.83032$ shows a marginal small variation 
$\Delta N /\langle N \rangle$ of 17\% in 5.0 yr (rest-frame) at a $3.7\sigma$ significance 
level (Table 3 and Fig. \ref{P1229MgII}). There is no variation of the associated four strongest 
Fe\,{\sc ii} lines ($1\sigma$: 7.3\%). The systems at $z_{\rm abs}=0.83102$ and 0.83128 are 
too weak to derive meaningful limits on their variabilty. There are two other Mg\,{\sc ii} systems 
at $\beta\leq 0.15$, and they are stable: the weak, single component doublet at $z_{\rm abs}=0.7684$ 
($1\sigma$: 8.7\%) and the triple system at $z_{\rm abs}=0.7564$ with one nearly saturated 
component ($1\sigma$: 20\%). 

Only one of the three targets with extremely strong saturated, low $\beta$ Mg\,{\sc ii} 
systems has proxy lines yielding meaningful limits. Toward PKS 0458$-$020, the Ni\,{\sc ii} 
triplet and the Zn\,{\sc ii}-Cr\,{\sc ii} lines at $z_{\rm abs}=2.03956$ are all of 
moderate strength and heavily blended without clear variation. In PKS 1741$-$038 spectra, the 
Fe\,{\sc ii} lines are saturated.  Thus, some variability constraint 
can only be placed on the $z_{\rm abs}=2.0545$ DLA absorber toward FBQS~J2340$-$0053: the 
three strongest S\,{\sc ii} components of medium strength  at  $z_{\rm abs}=2.0546$ 
and the strongest one of Al\,{\sc iii} at  $z_{\rm abs}=2.05511$ do not show variability 
($1\sigma$: 32\% and 28\%, respectively; see Sect. 4.2). 

%__________________________________________________%    	
%-----------------------------------------------------------	
%

\subsubsection{High $\beta$ Mg\,{\sc ii} systems}

For the high $\beta$ systems, variations are not detected from either their Mg\,{\sc ii} 
doublets or proxy lines. 
Toward AO 0235$+$164, the two strongest components of the Ca\,{\sc ii} doublet at $z_{\rm abs}=0.5239$, 
associated with the DLA, do not vary ($1\sigma$: 26\%). 
Toward PKS 0458$-$020, the Fe\,{\sc ii}$\lambda \lambda$2344,2382 lines 
at $z_{\rm abs}=1.5271$ (selecting the 3 
unblended components) do not vary either ($1\sigma$: 15\%). 
Finally, toward FBQS~J2340$-$0053, the highly multiple Fe\,{\sc ii} system at $z_{\rm abs}=1.3606$, 
associated with C\,{\sc i} and 21 cm absorptions, is stable ($1\sigma$: 14\% for all the 
components, using the weak Fe\,{\sc ii}$\lambda \lambda$2249,2260 doublet to constrain 
the other nearly saturated lines).

%section 4
\section{Discussion}

\subsection{Presence of intrinsic systems in blazar spectra ?} 

Time variability of narrow absorption lines at low $\beta$ have been securely detected in 
small samples of high ionization systems (involving C\,{\sc iv}, N\,{\sc v} and O\,{\sc vi}), 
observed at 
low and intermediate resolution (Narayanan et al. 2004; Wise et al. 2004, Hamann et al. 2011). 
The lines showing 
variations are essentially from mini-BALs and associated ($\Delta v <5000$ km s$^{-1}$) 
narrow-line systems. The time-scales investiged (in the quasar rest-frame) range from 
0.28 to 8.8 yr, and the variable narrow lines remained fixed in velocity. 
High and low ionization systems (at $\beta$ $>0.01$) with repeated observations have been 
identified by Hacker et al. (2013) in a very large sample of SDSS-DR7 quasar spectra: 33 out 
of 1084 systems have absorption lines with variations in their equivalent widths.  
For most of them $W_r>0.3$ \AA\ and $\Delta W_{r}/W_{r}>50$\%, with two-thirds 
of the variable systems being at 'relatively low' $\beta$ ($< 0.22$). 
 Low ionization species are present in all the 33 variable systems and C\,{\sc iv} 
is detected in the 17 systems at sufficiently high $z_{\rm abs}$ ($>1.50$). 
Chen et al. (2013) focussed their variability study on  associated ($\beta<0.03) $
Mg\,{\sc ii} systems identified in SDSS-DR7 and SDSS-DR9 (BOSS survey) spectra. They do not 
detect time variations  at a significance level larger than $3\sigma$ in their sample 
of 36 Mg\,{\sc ii} doublets. 

\begin{table*}[ht]
% [hb]  	
\caption[]{Targets with absorption systems showing marginal or significant variability.}	
\begin{center}
%\begin{tabular}{lcccccr}	
\begin{tabular}{l@{\hspace{3.5mm}}l@{\hspace{3.5mm}}l@{\hspace{3.5mm}}c@{\hspace{3.5mm}}c@{\hspace{3.5mm}}c}
\hline	
\noalign{\smallskip}	
  target &  $z_{\rm abs}^+$  & date & $N$ & date   &  $N$     \\	
  & &  & cm$^{-2}$ & & cm$^{-2}$    \\	
\noalign{\smallskip}	
\hline	
\noalign{\smallskip}
AO 0235$+$164  &  Mg\,{\sc ii} &(blend)  & marginal variability$^a$   &  &      \\
 & 0.85232  &  12-1997 & $(2.79 \pm 0.46) \times 10^{12}$   & 02-2007  &  $(2.82 \pm 0.38) \times 10^{12}$    \\
 & 0.85237{\boldmath $^{\rm v}$}  &  12-1997 & $(5.58 \pm 0.85) \times 10^{12}$   &  02-2007 &  $(3.25 \pm 0.57) \times 10^{12}$  \\
 & Mg\,{\sc ii} &  & no variability  &  &         \\
 & 0.85562  &  12-1997 &  $(7.27 \pm 0.36) \times 10^{11}$   &  02-2007  &  $(7.84 \pm 0.28) \times 10^{11}$  \\
% &  &  &  &  &         \\
\hline	
\noalign{\smallskip}
PKS 1229$-$02     & Na\,{\sc i} &   & significant variability$^b$ &  &        \\
 & -0.00013{\boldmath $^{\rm V}$} & 02-2002 &  $(1.98 \pm 0.41) \times 10^{11}$  & 04-2011 & $(4.6 \pm 1.5) \times 10^{10}$   \\
 & -0.00010 & 02-2002  & $(1.81 \pm 0.20) \times 10^{11}$  & 04-2011 & $(1.67 \pm 0.16) \times 10^{11}$   \\
 & Mg\,{\sc ii} &  & no variability  &  &         \\
 & 0.76836 &  02-2002 & $(7.15 \pm 0.24) \times 10^{11}$ & 04-2011 & $(7.57 \pm 0.58) \times 10^{11}$   \\
 & Mg\,{\sc ii} &  & marginal variability$^c$  &  &         \\
 & 0.83032{\boldmath $^{\rm v}$} &  02-2002 & $(3.42 \pm 0.07) \times 10^{12}$ & 04-2011 & $(2.87 \pm 0.13) \times 10^{12}$   \\
\hline	
\noalign{\smallskip}
FBQS~J2340$-$0053  & C\,{\sc i} &  (blend)  & marginal variability$^d$   &  &      \\
  & 2.05470 & 08-2006 & $(1.33 \pm 0.16) \times 10^{13}$  & 10-2008 & $(1.25 \pm 0.07) \times 10^{13}$  \\
  & 2.05473{\boldmath $^{\rm v}$} & 08-2006 & $(3.13 \pm 0.49) \times 10^{13}$  & 10-2008 & $( 1.52 \pm 0.23) \times 10^{13}$  \\
  & H$_2$,J=3 &    & marginal variability$^d$ &  &      \\
  & 2.05499 & 08-2006 & $(1.66 \pm 0.10) \times 10^{15}$  & 10-2008 & $(1.43 \pm 0.09) \times 10^{15}$  \\
  & 2.05513{\boldmath $^{\rm v}$} & 08-2006 & $(6.19 \pm 0.50) \times 10^{14}$  & 10-2008 & $(4.39 \pm 0.31) \times 10^{14}$  \\
\noalign{\smallskip}	
\hline	
\noalign{\smallskip}
\multicolumn{6}{l}{$^{a, b, c, d}$  \ See Fig. 4, 2, 5, 3 (left panel), 3 (right panel). } \\
\multicolumn{6}{l}{{\boldmath $^{\rm V}$} \ Significant variability: significance level $>$3.5$\sigma$ and 
$\Delta N /\langle N \rangle >$ 25\%.} \\
\multicolumn{6}{l}{{\boldmath $^{\rm v}$}  \ Marginal variability: significance level either 
in the range 2.0-3.5$\sigma$ or $>$3.5$\sigma$ but  $\Delta N /\langle N \rangle <$ 25\%.} \\
\end{tabular}	
\end{center}	
\label{obs}	
\end{table*} 

%-------------------------------------------------------------------------------------------

No variation of intervening Mg\,{\sc ii} systems in high spectral resolution spectra has been 
reported so far. Among the five targets in our sample, three blazars have in total six Mg\,{\sc ii} 
doublets at $\beta\leq 0.15$ which are suitable (unsaturated, not heavily blended lines) for 
variability studies. 
As described in Sect. 3.4, only marginal column density variations have been 
detected for two $z_{\rm abs} \sim 0.85$ Mg\,{\sc ii} absorbers at a significance level of 
2.3$\sigma$ and 3.7$\sigma$, respectively. The other four Mg\,{\sc ii} absorbers, at 
$z_{\rm abs} \sim 0.75$-0.85, are stable with $\Delta N /\langle N \rangle$ in the range 6-20\% 
at 1$\sigma$ confidence level.  Among them, the one in the spectrum of AO 0235$+$164 is 
150 km s$^{-1}$ away from the tentatively variable system.

In Bergeron et al. (2011), we found a potential excess of Mg\,{\sc ii} systems at 
$\beta \approx 0.1$ in blazar spectra and proposed that gas swept up by the AGN jet could 
be responsible for it. In such a scenario, absorption line variability can be expected, 
but, in the absence of predictions from numerical simulations, it is difficult to 
anticipate the time-scale and magnitude expected for the column density changes (this should 
strongly depends on the occurrence of marked variations in the activity of the nucleus itself 
at about the time when the blazar was observed). 
Thus, we cannot really draw conclusions from the lack of clear evidence 
for time variations among the limited set of systems which we could investigate here.

\subsection{ Structure in moderately ionized halo gas } 

\begin{figure}
   \centering
	\includegraphics[width=9.cm,angle=270]{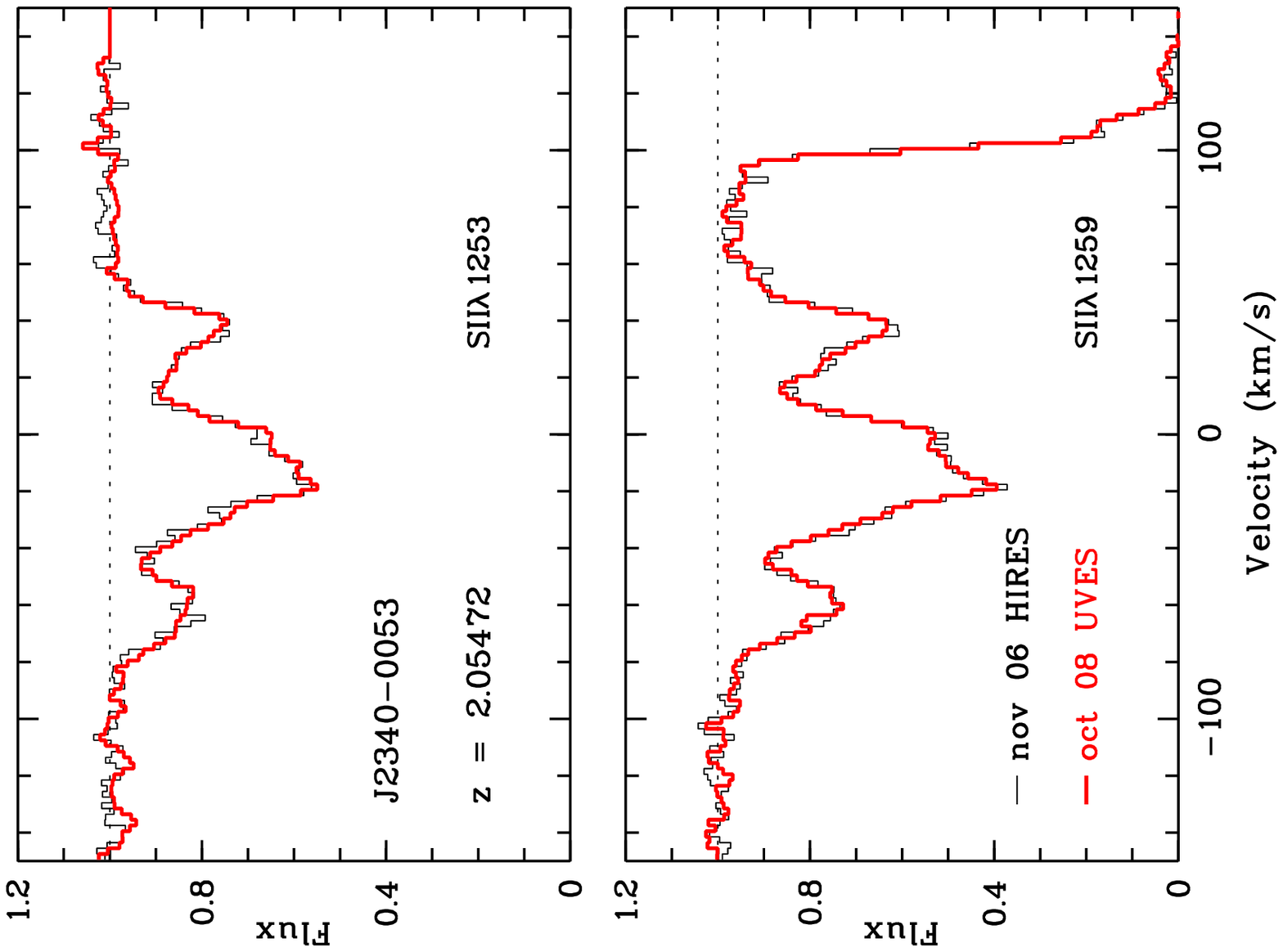}
    \caption{ S\,{\sc ii}$\lambda\lambda 1253,1259$  absorption
     at $z_{\rm abs} = 2.05454$ toward  FBQS~J2340$-$0053   
    in the november 2006 (HIRES, black line) and october 2008 (UVES, red line) spectra. }
         \label{SII}	
   \end{figure}

We now discuss systems involving narrow absorption lines associated with halo gas which is 
optically thick to UV ionizing radiation, i.e. gas classically traced by Mg\,{\sc ii} or 
Fe\,{\sc ii} transitions. We mentioned above the few cases for which we could get stringent 
upper limits on column density variations for discrete velocity components seen through intermediate 
opacity lines. One good example is the $z_{\rm abs} = 1.3606$ system toward FBQS~J2340$-$0053 
involving 10 Fe\,{\sc ii} components. Even when most of the absorption profile is optically 
thick  (this is often the case for the Mg\,{\sc ii} doublet), the data can be used to search 
for variations near the profile edges (e.g. appearance or disappearance of faint components). 
We could not identify 
any change of this kind. Moreover, since the component redshifts are let free when fitting the 
profiles, we could also get constraints on velocity variations; again, we found no significant 
change, with 3$\sigma$ upper limits generally below 1 km~s$^{-1}$. Occasionally, some line 
profiles are observed to be rather smooth as seen for the resolved S\,{\sc ii}$\lambda 1253$ 
and 1259 lines at $z_{\rm abs} = 2.05454$ in the FBQS~J2340$-$0053 spectrum. 
As shown in Fig. \ref{SII}, no significant difference is seen along the entire profile. 
The 1$\sigma$ upper limit on the relative flux variation is 1.7~10$^{-2}$ per resolution element
along the S\,{\sc ii}$\lambda 1259$ line profile. Since the latter spans 
over 120~km s$^{-1}$, the previous upper limit applies to about $120/6.6 \simeq 18$ 
distinct resolution elements.
%------------------------------------------------------------------------------------------%    	
%-----------------------------------------------------------	

To conclude for these systems, we detected no structure in the density and velocity field within 
Mg\,{\sc ii}/Fe\,{\sc ii} halos at the scales that we could probe (typically of the order of 100~au). 
This is consistent with previous observations 
of moderately ionised gas in galaxy halos. Indeed, both Rauch et al. (2002) and Kobayashi et al. 
(2002) find that a typical value for the cloud size associated with the individual velocity 
components is 200~pc, implying very little structure at the much smaller scales that we 
probe here. 
We also note that, 
from the analysis of the number density of strong Mg\,{\sc ii} systems detected on top of quasar 
emission lines in SDSS spectra, Lawther et al (2012) derive a lower limit of about 0.03~pc 
for the individual Mg\,{\sc ii} absorbing clouds. Finally, Welty (2007) analysed multi-epoch HST 
spectra of HD~219188 and detect no variations for S\,{\sc ii}, O\,{\sc i}, Si\,{\sc ii} and 
Fe\,{\sc ii} in the disk of our own Galaxy, over scales of the order of 100~au. Thus, our 
results are in good agreement with all known constraints on structure in moderately ionised gas. 

In this context, the variations tentatively detected by Hacker et al. (2013) for narrow 
intervening Mg\,{\sc ii} systems are most surprising. These authors analysed 
more than 1000 reliable quasar absorption systems with multi-epoch observations and 
extracted a small subset of 33 systems 
displaying some evidence for time changes in their equivalent widths. Their study
differs markedly from ours in the number of systems investigated and, given the small fraction 
of systems showing variability - about 3\% - it is true that we expect very few variable lines 
in our sample. However, it should be stressed that the two studies differ also strongly 
in the sensitivity expressed in terms of $\Delta W_r$ or $\Delta W_r/W_r$. The
variations reported in the SDSS spectra are very large with in most cases 
$\Delta W_r > 0.3$~\AA\ and $\Delta W_r/W_r > $ 50\%, corresponding to even larger 
fractional column density variations. For comparison, the marginal variation quoted 
above in the Mg\,{\sc ii}$\lambda 2796$ line at $z_{\rm abs} = 0.83032$ toward 
PKS~1229$-$02 is smaller than $\Delta W_r =0.01$~\AA\ and $\Delta W_r/W_r =5$\%. 

If the large variations quoted by Hacker et al. (2013) are real, one can imagine that 
the number of systems expected to display 
changes at the level that we can reach in our high resolution study should be much larger, 
coming in contradiction with the absence of clear detection of variations in our sample. 
% One way to reconcile the results from both studies could be to postulate the 
% existence of a specific population of rare absorbers, with distinct variability 
% properties.
Before drawing any conclusion, additional observations aiming at confirming 
the status of the 33 potentially variable systems would be highly desirable. In particular, 
since variations are (to first order) expected to be regular, a third spectrum should 
confirm the trend observed on the two first ones. Such a work is currently in progress
(see Appendix B2 in Dawson et al. 2013). Repeated {\it high resolution} observations of the 
brightest quasars involved would also be very useful to check the reality of time 
variations and to characterize them.

\subsection {High $z$ neutral/molecular gas}

C\,{\sc i} and H$_2$ are known to be closely related in diffuse molecular clouds
(Srianand et al. 2005). Indeed, the velocity components found for both species are generally 
similar, although some significant differences in their relative strength can be observed, as 
is the case for the $z_{\rm abs} = 2.0545$ system toward FBQS~J2340$-$0053 (Sect. 3.3 and 
Fig \ref{CIJ2340}).
We successively discuss these two tracers below.

C\,{\sc i} lines are detected in three high $z$ systems, one toward PKS~0458$-$020 and two 
toward FBQS~J2340$-$0053, but useful constraints could be obtained only for the latter two
at $z_{\rm abs} = 1.3606$ (no change in $N$) and 2.0454 (possible variation of about 70\% for 
an unresolved component). Unfortunately, since i) the transverse velocity of the quasar is
unknown and ii) the LoS drift is determined solely by the quasar motion for the  
$z_{\rm abs} = 2.0545$ system because $z_{\rm abs} \simeq z_{\rm em}$, it is not possible to 
compute the scale probed through the neutral gas. Further, the transverse observer's velocity 
for this target being only 12~km s$^{-1}$, little drift due to the observer's motion is expected
for the $z_{\rm abs}=1.3606$ absorber. We can then just get a typical value for the drifts by 
assuming a transverse velocity of about 300 km s$^{-1}$ for the intervening galaxies. This value 
together with a time interval of 1.9 yrs (observer's frame) between the two observations 
corresponds to a linear scale of about 50 and 40 au at $z_{\rm abs} = 1.3606$ and 2.0545 respectively. 
In our own Galaxy, the available information on structure in the C\,{\sc i} distribution remains 
quite limited. Welty (2007) performed a detailed study of the time behavior observed for visible 
and UV transitions due to an intermediate velocity cloud toward the star HD~219188. Variations 
by a factor of about two are seen for Na\,{\sc i}, C\,{\sc i} and Ca\,{\sc ii} column densities 
over scales of tens of au. Qualitatively, Na\,{\sc i} and C\,{\sc i} display the same time behavior,
indicating that the more extensive results obtained on the Na\,{\sc i} structure (Crawford, 2003; 
Smoker et al. 2011) are probably valid also for C\,{\sc i}. In this context, results obtained for 
$N$(C\,{\sc i}) at $z_{\rm abs} = 1.3606$ and 2.0545 are certainly consistent with the properties 
known for Galactic gas. Clearly, studies involving more systems are desirable to reach firm 
conclusions.

Regarding H$_2$, velocity components number 5 and 6 display an intermediate 
opacity which provides a good sensitivity for the $J=3$ column densities (the other $J=3$ 
components are either too weak or too saturated; similarly, $J=2$ lines are too strong
while $J=4$ are too weak to be suitable). 
The detection of marginal variations in component number 6 is 
interesting in the context of questions related to the "warm H$_2$" observed along many 
Galactic LoS (Gry et al. 2002; Jensen et al. 2010).
UV pumping is generally found to be insufficient to explain the observed excess of $J \ge 2$ H$_2$ 
with respect to the amount expected for a thermal distribution (Gry et al. 2002). 
Small structures (like shocks or turbulent vortices) where the temperature can locally reach 
high values have been proposed to account for both the production of species like CH$^+$ and the
relative amount of $J \ge 2$ H$_2$ (Godard et al. 2009). Presently, there are no direct 
constraints on the size of these putative regions. In our Galaxy, the 
only repeated measurements in the far-UV were performed toward the runaway O star, HD~34078 
(Boiss\'e et al. 2009). Stringent constraints were derived from the analysis of 
damped line profiles associated with $J=0$ lines, but no limit could be obtained for the $J = 2, 3, 
4$ levels because the corresponding lines are insensitive to small relative changes in $N$. 
It is noticeable that for the varying component toward FBQS~J2340$-$0053, the $N(J=0, 1, 2, 3)$ 
values (1.39 $\times 10^{17}$, 2.61 $\times 10^{17}$, 4.71 $\times 10^{15}$ and 
5.28 $\times 10^{14}$~cm$^{-2}$; for 
the $J=3$ column density, we give the average of the 2006 and 2008 values) are 
consistent with a unique excitation temperature, $T_{\rm ex} = 120$~K, unlike what is seen 
over most of Galactic LoS where an excitation temperature of about 300~K is required to fit 
$N(J)$ values at $J \ge 2$ (Jensen et al. 2010).
This indicates that the gas is dense enough to ensure thermalisation, in qualitative agreement 
with the small size suggested by the presence of time variations for this component. 

\subsection{ Local Galactic gas}

We found a large variation (factor of four) in Galatic $N$(Na\,{\sc i}) for one of the five 
velocity components toward PKS~1229$-$02. The constraint on the corresponding cloudlet size can 
be estimated from the peculiar motion of the Sun with respect to the LSR (see Sect. 2.2). For 
this direction, the Sun's velocity is mostly transverse with V$_t$ (Sun)/LSR = 17.8 km 
s$^{-1}$, implying a drift of about 35~au (this is only an estimate since we cannot know the 
transverse velocity of the cloud). This is well within the range (typically a few 10~au or 
larger) for which Na\,{\sc i} line changes have been observed on other LoS 
(Crawford 2003). The other four components toward PKS~1229$-$02 remained stable over the 
same linear scale, indicating that the associated fragments are characterised by sizes larger 
than 100~au. For other targets providing useful constraints on Galactic gas structure, we 
find variation neither toward AO 0235$+$164 nor toward FBQS~J2340$-$0053. The transverse velocity of 
the Sun for these two targets is 15.8~km~s$^{-1}$ and 16.6~km~s$^{-1}$ respectively, 
implying drifts of about 30 and 7~au in local material. In these two cases, 
Na\,{\sc i} lines have a relatively large opacity implying a poor sensitivity to variations.

Toward PKS~1741$-$038, the estimated drift over the 9.8~yr separating the two observations 
is about 17~au. The absence of variations in DIBs near 5780.5 and 5797.1~\AA\ can be related 
to results obtained recently by Cordiner et al. (2013) for $\rho$ Oph stars.
These authors find variations of about 5-9\% for the above DIBs over scales exceeding 
300~au. Thus, given our limited sensitivity, the observed stability is not surprising.

\section{Summary and prospects for further studies }

\noindent
As discussed above, the only clear and unambiguous variation has been detected 
for the Na\,{\sc i} doublet from our own Galaxy toward PKS~1229$-$02, in good agreement with 
previous results on the au-scale structure of neutral Galactic gas. In the same 
kind of media but at higher redshifts ($z \approx 2$ toward FBQS~J2340$-$0053), 
we also find tentative evidence for structure in neutral and molecular gas : 
C\,{\sc i} with $\Delta N /\langle N \rangle \simeq$ 70\% at the $3.0\sigma$ 
significance level and H$_2$ with $\Delta N /\langle N \rangle \simeq $ 35\%, 
at the $3.1\sigma$ level. For low $\beta$ absorption systems in blazar spectra, 
absorption profiles are, for the most part, very stable and only a marginal 
$3.7\sigma$ 17\% variation has been seen in the strength of a narrow Mg\,{\sc ii} 
component toward PKS~1229$-$02. On the other hand, no change was seen for moderately 
ionised gas at high $\beta$ (i.e. intervening halo gas), which again is consistent
with known constraints on typical cloud sizes in such media.

The results discussed above illustrate well the potential and the limits of the method 
used in this paper to investigate the small-scale structure of absorbing gas in 
quasar spectra. 
Let us first outline the great value of high spectral resolution. Often, several transitions
from the same species are detected and can be used to fit absorption line profiles, 
thus yielding well constrained $b$ and $N$ values for the velocity components seen 
at each epoch. This redundancy provides a powerful objective way (at least for those components 
which are partially resolved) to assess the reality of time changes and rule out false 
variations. Indeed, it is very unlikely to find by chance two artefacts mimicking 
true variations, i.e. occurring on 
two line profiles at the same position in velocity space and with a strength consistent 
with the $f$ values of the transitions involved. 
If species displaying more than two transitions (like C\,{\sc i} or H$_2$) can be used 
to increase the redundancy and the range covered by $f$ values, 
the probability is even lower and better constraints can be derived from line 
fitting. Conversely, the Mg\,{\sc ii} system at $z_{abs} = 0.85236$ toward AO~0235$+$164
illustrates well that with only two transitions differing in their $f$ values by a factor
no larger than two, one may be unable to reach firm conclusions when variations 
are moderate.
Furthermore, high resolution allows us to quantify the 
changes in meaningful astrophysical 
parameters, $N$ and $b$ values, or to set upper bounds on their variations. 
Limitations appear however for some line profiles (e.g. the C\,{\sc i} component 
toward FBQS~J2340$-$0053 for which variations are suspected), when the decomposition into 
discrete components required by line fitting is poorly defined or not unique. 
Such difficulties mean that the spectral resolution used is not high enough to analyse
properly the velocity distribution, implying that line opacities as well as $N$ 
and $b$ values cannot be determined reliably. One should therefore prefer absorption 
systems with simple velocity structure for repeated observations. 

In our study, we considered several topics for which useful constraints
can be obtained from an investigation of time variations. We now discuss
prospects for two of them. We mentioned above that generally, we have a 
poor knowledge of the scales probed by repeat observations because the 
transverse motion of the target is not measurable. The situation is more 
favorable for low redshift absorbers because the drift of the LoS 
through the intervening gas is determined mainly by the observer's motion, 
which is inferred directly from observations of the CMB dipole.
For instance, in order to study whether the small-scale structure seen in 
the Galactic neutral or weakly ionised gas
is ubiquitous in external galaxies, one could reobserve quasars belonging to 
the so-called quasar-galaxy pairs (see e.g. Bowen et al. 1991) and search for 
temporal changes in the Ca\,{\sc ii} or Na\,{\sc i} lines over more than 20 years. 
Of course, targets located in directions which are more or less normal to our 
large-scale peculiar velocity should be preferred (in this favorable 
configuration, the scales probed over 20 years are of the order of 1500~au).
Regarding high redshift absorbers, for which only a rough estimate of the scales probed 
can be obtained, one needs to observe a statistically significant sample of targets 
to derive meaningful results concerning the presence and strength of small-scale 
structure. Some quasars are found in pairs (Hennawi et al. 2010) and 
occasionally, absorbers appear to belong to galaxy clusters (Whiting et 
al. 2006) implying increased peculiar velocities. Such LoS will, in average, 
be associated with larger drifts, and should be targetted preferentially.

A second topic for which repeated observations of quasars at high spectral
resolution are especially timely involves the properties and origin of 
the warm ($J \geq 2$) H$_2$ found in diffuse molecular gas in our own 
Galaxy, as well as in high redshift absorbers. 
During the past ten years, the number of high $z$ H$_2$ systems 
has increased very significantly and now exceeds 20 (Albornoz-V\'asquez et 
al. 2014). Some of them display narrow velocity components 
which are well suited to a detailed study (see e.g. Noterdaeme et al. 2007).
As discussed in Sect. 3 for the system at $z_{\rm abs} = 2.0545$ toward 
FBQS~J2340$-$0053, multi-epoch observations can provide
key information on the typical size of regions where diffuse molecular gas is heated 
to temperatures above 200~K. Such high $z$ warm H$_2$ is seen at epochs when 
star formation was much larger than today (Madau \& Dickinson 2014) and one can therefore 
anticipate that the diffuse molecular medium should be strongly impacted by a very 
high dissipation rate of the mechanical energy carried by shock waves and turbulence. 

To conclude, we wish to stress that the data used here were 
acquired for purposes other than variability searches and thus, are certainly not optimal 
for our study. First, the two spectra available for each target often 
differ significantly in their S/N ratio and one is then limited
in the comparison by the lower quality spectrum. The time interval also varies a 
lot from target to target (unfortunately, the lag for the quasar displaying the 
richest spectrum - FBQS~J2340$-$0053 - is of only two years). 
A straightforward strategy to get better constraints or address some 
questions discussed above with more appropriate data would consist in i) selecting 
targets displaying the required types of absorption lines and with one already available 
good high resolution spectrum taken at least 10 years ago, and ii) reobserving them at 
a similar S/N and resolution. Obviously, the existence of databases where high 
resolution quasar spectra are archived properly are extremely useful in such an approach.  
Finally, we note that improvements in the accuracy of wavelength calibrations 
accomplished recently in the context of the ESO Large Program designed to constrain the 
variation of fundamental constants (Molaro et al. 2013) or forseen for instruments that 
will become available in the near future (e.g. the {\it ESPRESSO} spectrograph to be 
mounted on the ESO/VLT telescope) should indeed help to investigate in more detail 
the velocity field within the absorbers.\\

\begin{acknowledgements}
We are very grateful to several colleagues who provided some of the
spectra used in our study or contributed to their reduction: John Black
(for the 2001 PKS~1741-038 spectrum), C\'edric Ledoux (PKS~0458-020, UVES data), 
Pasquier Noterdaeme and Hadi Rahmani
(FBQS~J2340-0053, UVES data), John O'Meara (AO~0235+164, HIRES spectrum)
and Giovanni Vladilo (UVES spectra of AO~0235+164 and PKS~1229-02). Thanks
are also due to Jean-Philippe Uzan for discussions about the estimate of
line of sight drifts on cosmological scales and to Gary Mamon for help on
data handling techniques. We are grateful to the referee for several constructive 
comments that helped to clarify the content of this paper.
\end{acknowledgements}

\end{document}